\documentclass[twocolumn]{aastex631}

\usepackage{natbib,hyperref}

\usepackage{graphics,graphicx}
\hypersetup{urlcolor=blue}
\usepackage{natbib}
\usepackage[outdir=./]{epstopdf}
\usepackage{textcomp,gensymb}
\usepackage{xfrac}
\usepackage{xcolor}
\usepackage[utf8]{inputenc}
\usepackage[version=4]{mhchem}

\newcommand{\vsini}{$v\sin{i_*}$}

\newcommand{\kepler}{{\it Kepler}}

\newcommand\kms{km~s$^{-1}$}

\newcommand{\tess}{\textit{TESS}}
\newcommand{\ktwo}{{\textit K2}}

\begin{document}

\title{Ever Elusive Exospheres: One Probable Detection and Two Non-Detections of H$\alpha$ Transits in Young Systems}

\author[0000-0002-1312-3590]{Reilly P. Milburn}
\affiliation{Department of Physics and Astronomy, The University of North Carolina at Chapel Hill, Chapel Hill, NC 27599, USA}

\author[0000-0003-3654-1602]{Andrew W. Mann}
\affiliation{Department of Physics and Astronomy, The University of North Carolina at Chapel Hill, Chapel Hill, NC 27599, USA}

\author[0000-0003-1337-723X]{Keighley Rockcliffe} 
\affiliation{University of Maryland Baltimore County, Baltimore, MD, USA}
\affiliation{NASA Goddard Space Flight Center, Greenbelt, MD, USA}

\author[0000-0001-8045-1765]{Erin E. Flowers}
\altaffiliation{NSF Graduate Research Fellow}
\affiliation{Department of Astrophysical Sciences, Princeton University, 4 Ivy Lane, Princeton, NJ 08540, USA}

\author[0000-0002-8091-7526]{Alexis Heitzmann}
\affiliation{Space Research Institute, Austrian Academy of Sciences, Schmiedlstrasse 6, A-8042 Graz, Austria}
\affiliation{Institute for Theoretical Physics and Computational Physics, Graz University of Technology, Petersgasse 16, 8010 Graz, Austria}

\author[0000-0001-7516-8308]{Benjamin T. Montet}
\affiliation{School of Physics, The University of New South Wales, Sydney NSW 2052, Australia}

\author[0000-0002-4891-3517]{George Zhou}
\affiliation{School of Mathematics, Physics and Computing, The University of Southern Queensland, Toowoomba, 487-535 West St, Darling Heights QLD 4350, Australia}

\author[0000-0002-8399-472X]{Madyson G. Barber}
\altaffiliation{NSF Graduate Research Fellow}
\affiliation{Department of Physics and Astronomy, The University of North Carolina at Chapel Hill, Chapel Hill, NC 27599, USA} 

\begin{abstract}

Gaps in the exoplanet population, such as the Neptunian Desert, point to the importance of mass-loss in sculpting the radii of close-in exoplanets. Young planets ($<500$\,Myr) offer the opportunity to detect such mass-loss while it is still strong, and to test models of the underlying physical processes. We search for evidence of an H$\alpha$ transit in high-resolution spectra of three young planets, HD\,63433\,b (400\,Myr), DS\,Tuc\,A\,b (45\,Myr), and HIP\,67522\,b (17\,Myr) using HARPS-N, Magellan-PFS, and CHIRON respectively. We validate our method by testing it on several photospheric lines less impacted by stellar variability. We find no evidence of a transit signal for HD\,63433 b and  DS Tuc A b (3$\sigma$ limits of 0.9\% and 0.3\%, respectively). For HIP\,67522\,b, we detect significant excess absorption (3.44$\pm$0.28\%) aligned with the transit time, narrow compared to the stellar line, and blueshifted from the stellar rest frame. In combination, these suggest the signal is from the planet. However, stellar variation in the H$\alpha$ line over the course of the observations is comparable in size to the transit signature and the duration is shorter than the photometric transit, so this detection requires confirmation. Our findings, and other recent results in the literature, suggest that planets younger than 50 Myr are more favorable for the detection of atmospheric escape with H$\alpha$ observations, though older populations might still show escape in other diagnostics.
\end{abstract}


\section{Introduction} \label{sec:intro}

The success of NASA's \kepler{} \citep{Borucki2010}, \ktwo{} \citep{VanCleve2016}, and \tess{} \citep{2014SPIE.9143E..20R} missions has enabled astronomers to detect and confirm thousands of planets across a wide range of parameter space. This abundance of transiting exoplanets allows us to explore science questions like how planet occurrence varies with stellar type \citep[e.g.,][]{Mulders2015,2019AJ....158...75H}, orbital period \citep[e.g.,][]{Fressin:2013qy, 2016ApJ...828...99C}, planet size \citep[e.g.,][]{Dressing2013, Winn2015}, and the presence of a stellar companion \citep[e.g.,][]{Kraus2016a, Dupuy2022}.

One such discovery has been the role of mass-loss in sculpting the distribution of planetary systems. There is a notable scarcity of Neptune-radius planets at close orbital periods \citep[the hot Neptune Desert;][]{mazeh_dearth_2016-2}. This ``Neptunian desert'' provides strong evidence that sub-Saturns undergo significant atmospheric mass loss compared to their gas giant counterparts \citep{Lundkvist2016}. Hot Neptunes typically have lower gravitational potentials than their Jovian counterparts, making them more susceptible to photoevaporation and/or Roche lobe overflow \citep{Kurokawa2014}. Models of photoevaporation predict that low-density gaseous atmospheres should evaporate on relatively short timescales \citep{owen_evaporation_2017}, while Hot Jupiters are expected to lose only 1\% of their total mass \citep{adams_magnetically_2011}.

Mass-loss is also invoked to explain the bimodality in the radii of small (1.5$R_\oplus$-2$R_\oplus$), close-in ($P\lesssim$30\,days) planets \citep{OwenWu2013, Fulton2017}, sometimes called the small radius gap. In this scenario, lower-mass and/or more highly irradiated sub-Neptunes evolve into super-Earths. The major drivers of mass loss in this regime are thought to be high-energy radiation from the young host star \citep{Lopez2013, Chen2016}, or by the combination of stellar heating with remnant thermal energy from the formation of the planet \citep[core-powered mass-loss;][]{Gupta2019, Gupta2020}. This deficit of planets can also be explained if small planets form without an H/He atmosphere and some accrete a thin atmosphere \citep[a few percent of the total mass;][]{neil_joint_2020,Rogers2021}.

Studying planets at young ages can reveal the mechanisms responsible for the drastic changes in physical and orbital properties that can occur early in their lifetimes. For example, if the small radius gap is a product of formation, it should be visible as soon as the planet can be detected. Photoevaporation is generally thought to be strongest from the dispersion of the proto-planetary disc out to 100\,Myr, when the star is most active in the X-ray and EUV \citep{Jackson2012,Ansdell2015} and the planet is inflated to a low density \citep{mann_tess_2022,Owen2020}. However, recent work has suggested EUV-driven escape can continue on $\simeq$Gyr timescales \citep{King2021}. Core-powered mass-loss works on timescales of $\sim$Gyr \citep{Gupta2020}, and, if dominant, predicts no significant radius gap below a few hundred Myr.

Young planets appear to be inflated even out to 700\,Myr \citep{Mann2017a,Fernandes2022}, with one such planet at this age (K2-100b) found within the Neptunian desert \citep{2019MNRAS.490..698B}. Further, 0.1-1Gyr planets show similar rates of mass-loss compared to their older counterparts \citep{feinstein_h-alpha_2021, orell-miquel_mopys_2024}. This appears to favor processes that operate on longer timescales like core-powered mass-loss. However, the number of systems with measurements below 100\,Myr is quite small, despite the fact that this is the most critical period for photoevaporative mass-loss.

Testing for atmospheric escape is typically carried out through transit observations at wavelengths corresponding to maximum opacity of the escaping material (mostly H/He). The three most common tracers are Ly$\alpha$ in the ultraviolet \citep[e.g.,][]{Rockcliffe2023}, H$\alpha$ in the optical \citep[e.g.][]{yan_extended_2018}, and the He~I triplet in the near-infrared \citep[e.g.][]{Oklopcic2018}. Ly$\alpha$ is often the strongest, but observations are restricted to space-based facilities (mainly the \textit{Hubble Space Telescope}) and stars within 50~pc due to interstellar extinction. The He I triplet requires strong stellar X-ray and UV radiation to populate the metastable triplet state \citep{Oklopcic2018} \textit{and} low FUV radiation to avoid depopulating it \citep{Biassoni2024}. Young stars also have variable He lines \citep{Krolikowski2024}. 

The H$\alpha$ line has been less favorable because of both strong variability and because the planetary signal is expected to be far lower there than in Ly$\alpha$, sometimes yielding detections in the latter with none in the former \citep[e.g.][]{Ehrenreich2015,cauley_decade_2017}. The other problems for H$\alpha$ are the need for electrons populating the first excited state and the ionization fraction. However, \citet{Allan2019} showed that for young Jovian planets, $N_2/N_1$ is expected to be $10^{-5}$, which is both higher than for older equivalent planets and also sufficient for a transit depth of several percent in H$\alpha$. They also found the ionization fraction to be small within 1-5$R_J$ of the planet. While \citet{Allan2019} only considered hot Jupiters, many young planets like HIP\,67522\,b are low-gravity analogues of these systems \citep{Thao2024b}. Together, this suggests H$\alpha$ is {\it more favorable than Ly$\alpha$ at ages below 1\,Gyr.} However, recent work has found that the detection of H$\alpha$ absorption is no more likely in 0.1-1\,Gyr planets than their older counterparts \citep{Orell-Miquel2024}. It may be that $<$100\,Myr planets provide the best targets for H$\alpha$.

In this work, we present results for three young planetary systems using H$\alpha$ as a tracer for atmospheric escape. Specifically, we focus on HD 63433b (400 Myr), DS Tuc Ab (45 Myr), and HIP 67522b (17 Myr), which have optical spectra taken for measuring the planet's orbital alignment with respect to the host star's rotation (Rossiter-McLaughlin or Doppler tomography). This data covers the H$\alpha$ line and was taken at high SNR needed to detect a planetary transit, but requires special processing to remove tellurics and fit for stellar variations.

\section{Targets, Observations, and Reduction} \label{sec:obs}

\begin{figure}[t]
\centering
\includegraphics[width=0.49\textwidth]{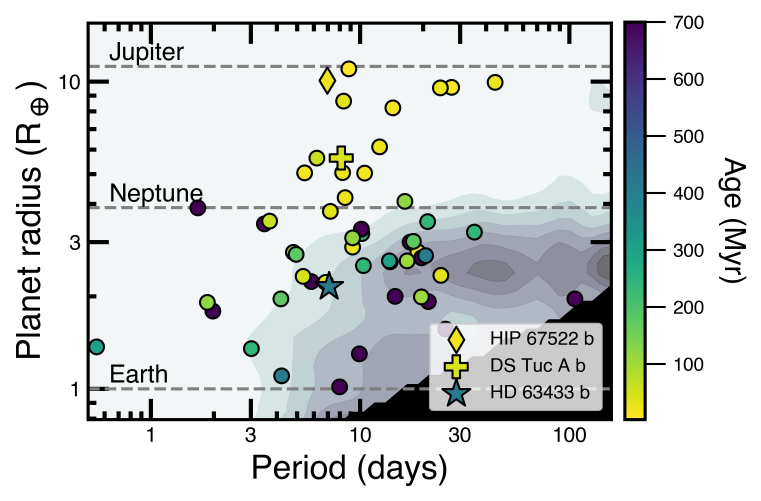} 
\caption{The young planet population in period-radius space. The color denotes the age of the planet. HIP 67522 b is marked by a diamond, DS Tuc A b by a cross, and HD 63433 b by a star. The contours indicate the relative occurrence of planets inferred from the  \kepler\ distribution \citep{Dattilo2023}. The young planet population shown is not corrected for completeness, although young planets are known to be larger than their older counterparts even after accounting detection biases \citep{Vach2024}.} 
\label{fig:sample}
\end{figure}

The three planets examined here were selected because they are young ($\ll 1$~Gyr: Figure~\ref{fig:sample}) and have prior high-resolution optical spectroscopy taken in transit. The existing data sets were taken to measure the spin-orbit alignment of the planets via Rossiter-Mclaughlin or Doppler tomography. 

The parameters for each planet used in this analysis were taken from the relevant discovery paper. We provide the basic planet parameters used in this work and the relevant reference Table~\ref{tab:planet}.

\begin{deluxetable*}{lccccc}
\tablewidth{10pt}
\tablecaption{Planetary Properties} {\label{tab:planet}
}
\tablehead{
\colhead{} &
\colhead{\textbf{HIP67522 b}} & \colhead{\textbf{HD63433 b}} & \colhead{\textbf{DS Tuc A b }}}
\startdata
\hline
Planet Radius ($R_\oplus$) & $9.99^{+0.24} _{-0.22}$ & $2.112^{+0.093}_{-0.086}$ & 5.70 $\pm$ 0.17\\
\hline
Orbital Period (days) & 6.9594731 $\pm$ 0.0000022 & 7.1079379 $\pm$ 0.0000054 & 8.138268 $\pm$ 0.000011 \\
\hline
Age (Myr) & 17 & 400 & 45 \\
\hline
Transit Duration (hours) & $4.85^{+1.13}_{-0.36}$ & $2.93 \pm 0.29$ & $3.1764^{+0.0118}_{-0.0094}$ \\
\hline
Parameter Reference & \citealp[]{Barber2024a} & \citet{Capistrant2024} & \citealp[]{newton_tess_2019} & \\
\hline
Data Reference & \cite{heitzmann_obliquity_2021-1} & \citet{mann_tess_2020-3} & \cite{zhou_well-aligned_2020} \\
\hline
\hline
Mass ($M_\oplus$) & $13.8 \pm 1.0$ & $37.3 \pm 9.6$ & $<14.4$ \\
Planet Mass Reference & \cite{2024AJ....168..297T} & \cite{2024ApJS..272...32P} & \cite{2021AA...650A..66B}\\
\hline
Surface Gravity (m s$^{-2}$) & 1.35 & 78.1 & $<$ 4.34 \\
\hline
XUV irradiation (erg s$^{-1}$ cm$^{-2}$) & $333000$ & $4960$ & $54500$ \\
\hline
Energy-limited mass loss rate ($\times 10^{10}$ g s$^{-1}$) & $496$ & $0.0276$ & $<14.4$ \\
\enddata
\end{deluxetable*}

Significantly more details on the observations and data reduction can be found in the original papers: \citet{mann_tess_2020-3} for HD\,63433\,b, \cite{zhou_well-aligned_2020} for DS Tuc A\,b, and \cite{heitzmann_obliquity_2021-1} for HIP\,67522\,b. To briefly summarize, the observations covered two transits of DS Tuc A b, and one transit each for HIP67522 b and HD63433 b. All observations were taken under clear or mostly clear conditions. For signal-to-noise ratio (SNR) estimates, only the HD\,63433 dataset provided uncertainty estimates. For the others, we estimated the spectral uncertainties from the square root of the flux counts per pixel near the H$\alpha$ line. Additional details on the instrument, estimated SNR, and date of observations are given in Table~\ref{tab:obs}. 

\begin{deluxetable*}{lccccc}
\tablewidth{0pt}
\tablecaption{Summary of Transit Observations\label{tab:obs}
}
\tablehead{
\colhead{} &
\colhead{\textbf{HIP67522 b}} & \colhead{\textbf{HD63433 b}} & \colhead{\textbf{DS Tuc A b (1)}} & \colhead{\textbf{DS Tuc A b (2)}} 
}
\startdata
\hline
Polynomial fit order & 3 & 3 & 3 & 3 \\
\hline 
Region of integration (\AA) & 6562.0 - 6563.6 & 6562 - 6564 & 6561.46 - 6564.74 & 6561.46 - 6564.74  \\
\hline
\hline
Instrument & CHIRON & HARPS-N & Magellan-PFS & Magellan-PFS \\
\hline 
Instrument Resolution (R) & 80,000 & 115,000 & 130,000 & 130,000 \\
\hline
Observation Date (UT) & 2021-05-14 & 2020-03-07 & 2019-08-19 & 2019-10-07 \\
\hline
Total Observations & 24 & 43 & 36 & 33 \\
\hline
Number of In-Transit Spectra & 14 & 24 & 17 & 18 \\ 
\hline
Exposure Time (s) & 1200 & 420 & 600 & 600 \\
\hline
Typical SNR\tablenotemark{a} & 52 & 96 & 43 & 52 \\
\hline
\enddata
\tablenotetext{a}{SNR corresponds to the peak value in the order containing H$\alpha$ per pixel. SNR varies during the transit, so the average value is quoted. }
\end{deluxetable*}

\subsection{Planet X-ray/EUV flux and mass-loss rates}

We gathered X-ray ($0 - 100$ \AA) and EUV ($100 - 912$ \AA) measurements and reconstructions for all three host stars in order to estimate the mass loss rates of their planets. \cite{2024AJ....168..297T} included a panchromatic spectrum for HIP 67522, which incorporated a combination of Chandra measurements and APEC modeling for the X-ray and the differential emission measure technique for reconstructing the EUV (see their Section 4). \cite{2022AJ....163...68Z} included a similar panchromatic spectrum for HD 63433. We used the X-ray portion of that spectrum which came from Chandra observations and combined it with an updated EUV - X-ray luminosity relation from \cite{2025A&A...693A.285S} to get the EUV luminosity for HD 63433. \cite{2025ApJ...980...27K} published many Chandra observations of DS Tuc A in both quiescence and flare activity. We took the average of the quiescent measurements in the 0.5 - 10.0 keV band, which \cite{2025ApJ...980...27K} indicate is encapsulates the majority of DS Tuc A's X-ray output. We combined this with the differential emission measure EUV estimate for DS Tuc A provided by \cite{2025arXiv250715953F}. 

Table~\ref{tab:planet} lists the XUV flux received by the three planets in this work. A recent analysis from \citet{Maggio2024} obtained a similar XUV flux ($3.9\times10^5$erg s$^{-1}$ cm$^{-2}$), suggesting our estimate is reliable at the 10-20\% level. A larger effect than the formal uncertainties is variation in the frequency and magnitude of activity for young stars, which can change these estimates at the order-of-magnitude level. Such variation may change the mass-loss regime experienced by the irradiated planets, causing some to enter recombination- or photon-limited mass loss \citep{2016ApJ...816...34O}. It also illustrates the importance for characterizing the high-energy radiation of young planet hosts both in quiescence and short- and long-term activity \citep{Maggio2024}.

The escape rate for each planet is related to the depth of its gravitational potential well. We compiled the mass measurements and estimates for each of the planets in this work, which we list in Table~\ref{tab:planet} as well as the relevant references. We note that two of these (HD\,63433\,b and DS Tuc\,b) are based on radial velocities (challenging for young stars), while the mass of HIP\,67522\,b is inferred from its transmission spectrum \citep{Thao2024b}.

The energy-limited mass loss rate (Equation~\ref{eq:loss}) is formulated from directly translating the energy heating the upper atmosphere (the XUV irradiation; $F_\text{XUV}$) into energy required to escape the planet's gravity ($R^3_{\text{p}}/GM_{\text{p}}$). This formulation, reviewed in \cite{2019AREPS..47...67O}, is based on outflow measurements for the young Earth and Venus \citep{1981Icar...48..150W}. The $eta$ factor in Equation~\ref{eq:loss} accounts for the range of heating efficiencies within planetary atmospheres. We assumed $\eta = 0.1$ for our estimates, however it has been shown to range from $0.1$ to $0.6$ across different planets \citep{2014ApJ...795..132S}. The efficiency, along with the density and XUV irradiation of the planet, can govern which mass loss regime these planets are in \citep{2016ApJ...816...34O}. In order to be considered ``lost'', a planet's escaping atmosphere only needs to reach its Roche lobe. This is corrected for within the $K_{\text{eff}}$ factor, which includes the semi-major axis of the planet ($a$) and its radius ($R_{\text{p}}$).

\begin{equation} \label{eq:loss}
    \dot{M} = \eta \frac{\pi R^3_{\text{p}}F_{\text{XUV}}}{GM_{\text{p}}K_{\text{eff}}}
\end{equation}
\begin{equation} \label{eq:Keff}
    K_{\text{eff}} = \frac{(a/R_{\text{p}} -1)^2 (2a/R_{\text{p}} +1)}{2(a/R_{\text{p}})^3}
\end{equation}

The resulting mass-loss rates are given in Table~\ref{tab:planet}. HIP 67522 b, as the youngest and lowest-gravity planet exhibits the highest mass loss rate. The mass loss rate for HIP 67522 b and upper limit for DS Tuc A b are larger than similarly estimated loss rates for planets exhibiting escape ($\sim10^{10}$ g s$^{-1}$ for AU Mic b and GJ 3470 b; \citealt{2023AJ....166...77R, 2018A&A...620A.147B}).  This does not change in magnitude if we assume the planet is in the recombination-limited mass loss regime \citep{2008PhDT........25M}. It follows that a larger amount of hydrogen will populate the n = 2 energy state needed to absorb H$\alpha$ in transmission. 

\citet{Maggio2024} report a mass-loss rate for HIP\,67522\,b of $2.5\times10^{12}$g\,s$^{-1}$, about half our estimate. Most of this difference is due to a higher assumed planet mass (57$M_\oplus$ versus 14$M_\oplus$). Repeating their approach, as outlined in \citet{Caldiroli2022}, we get a mass-loss rate of $3.7\times10^{12}$g\,s$^{-1}$, in between the two estimates. The variation between these and the nature of the assumptions discussed above suggest uncertainties in this calculation at the factor of a few level for HIP\,67522\,b. The input data is worse for the other two, suggesting order-of-magnitude errors, but this is sufficient to explain the differences between the three systems and compare to other planets with similar escape data. Better mass-loss rate constraints could be placed on these planets using updated hydrodynamic models (e.g., {\tt p-winds}; \citealt{2022A&A...659A..62D}) that incorporate tracking the n = 2 state of hydrogen to compare to H$\alpha$ observations.

\section{Methods} \label{sec:methods}

Our approach roughly followed that of \citet{yan_extended_2018}'s analysis of the H$\alpha$ transit of KELT-9b. In addition to small changes based on the instrument and setup, we differed in our approach to removing tellurics from the spectra.

Our process began with the reduced time-series spectra. From there, the basic steps were:
\\
\begin{enumerate}
    \item Shift and fill spectra into the rest frame to account for Rossiter-McLaughlin shifts. 
    \item Continuum normalize and remove telluric features from the spectral order containing H$\alpha$ using the Python package TelFit \citep{gullikson_telfit_2014}.
    \item Median combine out-of-transit spectra to generate a template of the stellar signal free of transit signal. 
    \item Divide all spectra by the out-of-transit stellar template. The resulting spectrum should contain a mix of the planet, time-dependent stellar variations, and random noise.
    \item Visually inspect the time-series spectra to search for shifted absorption features \citep{Lavie2017,Allart2018}. 
    \item Measure the equivalent width of any H$\alpha$ feature in the spectrum, yielding a time-series we can inspect for evidence of a transit. 
\end{enumerate}
We describe each step in more detail below.

\subsection{Rossiter-McLaughlin Effect}\label{sec:rm_correction}

As explained in \citet{Casasayas-Barris2021}, the Rossiter-McLaughlin can distort the stellar line profile by blocking out preferentially blue- or red-shifted regions of the limb-darkened star. This can show up as excess absorption like that of a planet, except that the resulting signals covers the full line (while the planetary signal should be narrow). For rotationally broadened stars like those considered here, the RM-induced signal will be much broader than a planetary signal. For that reason we do not expect a significant impact, and should be able to easily separate RM-induced signals from those generated by excess planetary absorption. 

\citet{Casasayas-Barris2021} correct for this by filling in the line with a stellar model. This is challenging for our stars because the H$\alpha$ line varies, models do not reproduce this line as well as for older stars, and the correction factor depends on assumptions about things like microturbulence versus rotational broadening (the latter is small in older stars). Instead, we applied a basic correction using the best-fit parameters of the RM model from the original source (Table~\ref{tab:obs}). We shifted the spectrum and added in the missing flux assuming a Gaussian 'notch'. We assumed the width is a the instrument profile centered at the predicted velocity, with depth $\left(R_p/R_*\right)^2$. The correction was small compared to both random noise and H$\alpha$ variability, and the resulting planet spectrum underwent only a nominal change, which further confirms that this is a relatively small effect compared to random and systematic noise in the spectra and more sophisticated modeling is not required.

\subsection{Continuum Normalization and Telluric Removal}\label{sec:cont_norm}

Telluric features at these wavelengths are primarily due to absorption from water and oxygen species present in the Earth’s atmosphere. These features vary by location of observations and environmental conditions at the time of observing (e.g., the temperature, humidity, pressure, and airmass of the target). Some telluric features fall within the same region as H$\alpha$ and the nearby continuum region, which has the potential to affect the line profile and depth at levels comparable to that we expect from the planet. 

To remove tellurics, we used the \texttt{TelFit} package \citep{gullikson_correcting_2014}. For the telluric line model, \texttt{TelFit} uses the Line By Line Radiative Transfer Model (LBLRTM; \citealp[]{clough_lblrtm_2014}). LBLRTM accounts for the temperature, pressure, and atmospheric abundances, which are fit and scaled appropriately for each atmosphere layer. For this work, we fit only for the \ce{H2O} species, as water had the only significant telluric features near H$\alpha$. 

TelFit also handles continuum normalization. By default, TelFit uses a 7th order polynomial fit to the continuum. We compared 2nd- through 8th-order polynomials. The biggest differences were at the edges of the order, far from the H$\alpha$ line and the continuum region. High-order polynomials ($>5$) would often fit into the H$\alpha$ line, especially for HIP~67522. Ultimately, we chose a 3rd order polynomial fit for all observations. The best fit was chosen to minimize the fit order while taking into account the flatness of the continuum near the H$\alpha$ profile. 

We did not apply telluric removal with TelFit on HD63433 b. This data already received a telluric correction as part of the existing reduction, and the additional TelFit model introduced erroneous corrections and excess noise in our measurements. For consistency, we still modeled the continuum using TelFit.

TelFit also carries out a wavelength solution on the data using a 3rd order polynomial fit. We used this default wavelength solution as is without adjustments. 

\subsection{Stellar Template Generation and Removal} \label{sec:star_temp}

\begin{figure}[t]
\centering
\includegraphics[width=0.49\textwidth]{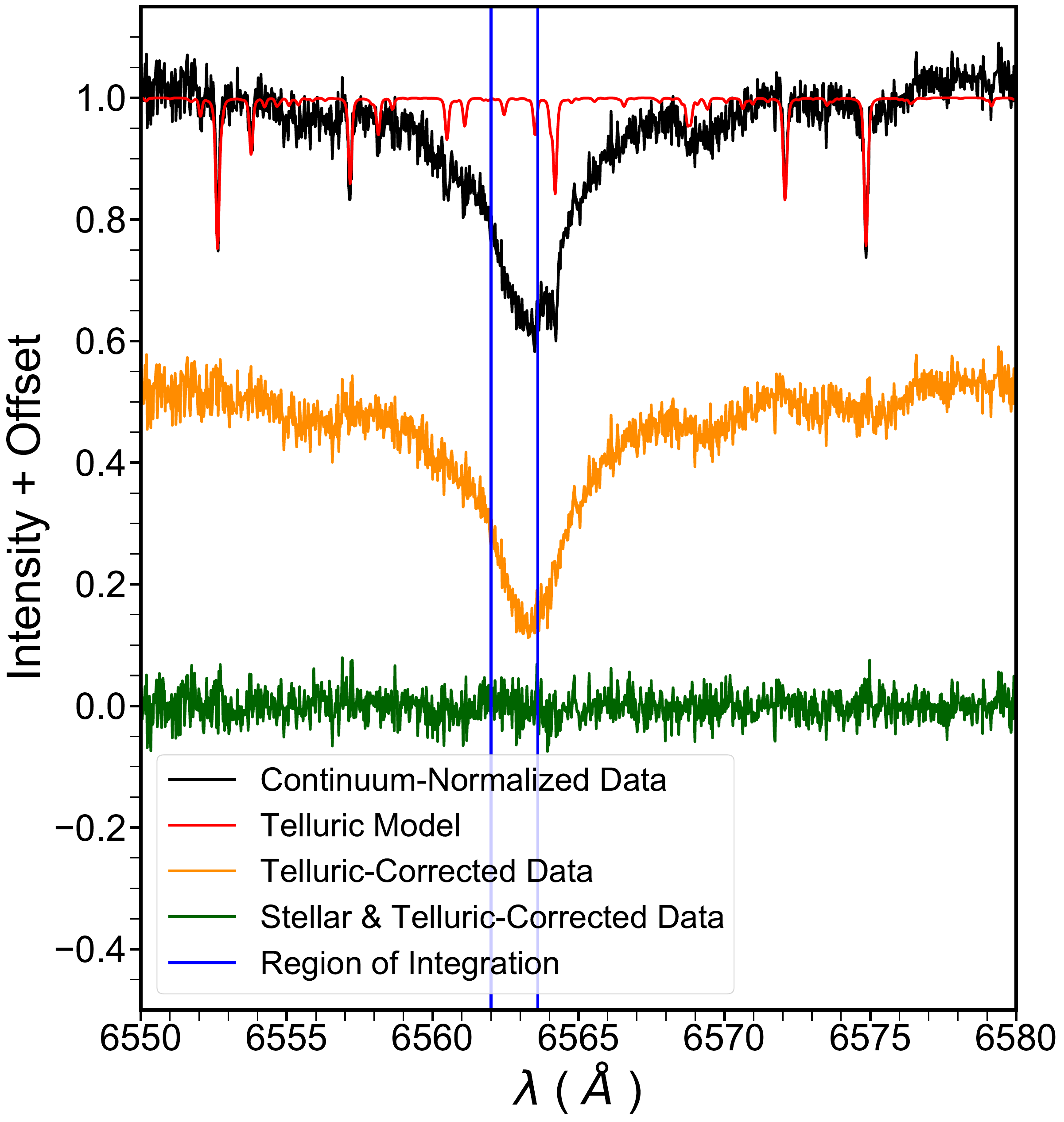}
\caption{A schematic depicting the data reduction process using a sample spectrum from observations of HIP67522 b. We represent the continuum-normalized raw data in black, the telluric model in red, the telluric-corrected data in orange, and the continuum-normalized telluric and stellar-corrected data in green. All steps in the reduction process depicted are offset by a scalar for better visualization. The region of integration used for HIP\,67522 b is noted by the vertical blue lines.}
\label{fig:reduction}
\end{figure}

To remove the signature of the star, we divided each continuum-normalized and telluric-removed spectrum by a template of the stellar H$\alpha$ profile. The template was generated by median combining the continuum-normalized and telluric-removed out-of-transit spectra. This included all observations before and after the predicted transit start and end times. We present an example of one continuum-normalized telluric-corrected and stellar-corrected spectrum in green in Figure \ref{fig:reduction}. 

Prior studies have found H$\alpha$ of young planet hosts varies by 2-50mA over the rotation period \citep[e.g.,][]{feinstein_h-alpha_2021,Damasso2023}. However, scaled to the transit window this is expected to be weaker than the signal from the escaping atmosphere. Instead, short-term variations would be dominated by flares. As a result, some variations between pre- and post- transit observations are expected, but we expect such variations to be unconnected to the transit timing and exhibit red/blue shifts consistent with the stellar surface.

\subsection{Examination of time-series spectra} \label{sec:}
If absorption due to an extended atmosphere is present, we expect it to be narrower than the stellar lines, as the planet has smaller rotational broadening (\vsini). Excess absorption should also be blue-shifted with respect to the star, as the outflow pushes towards the observer by the stellar wind \citep[although in rare cases it is redshifted, e.g.,][]{Lavie2017}. Thus, the first step is to identify the ideal region over which to calculate the equivalent width.

\begin{figure}[t]
\centering
\includegraphics[width=0.48\textwidth]{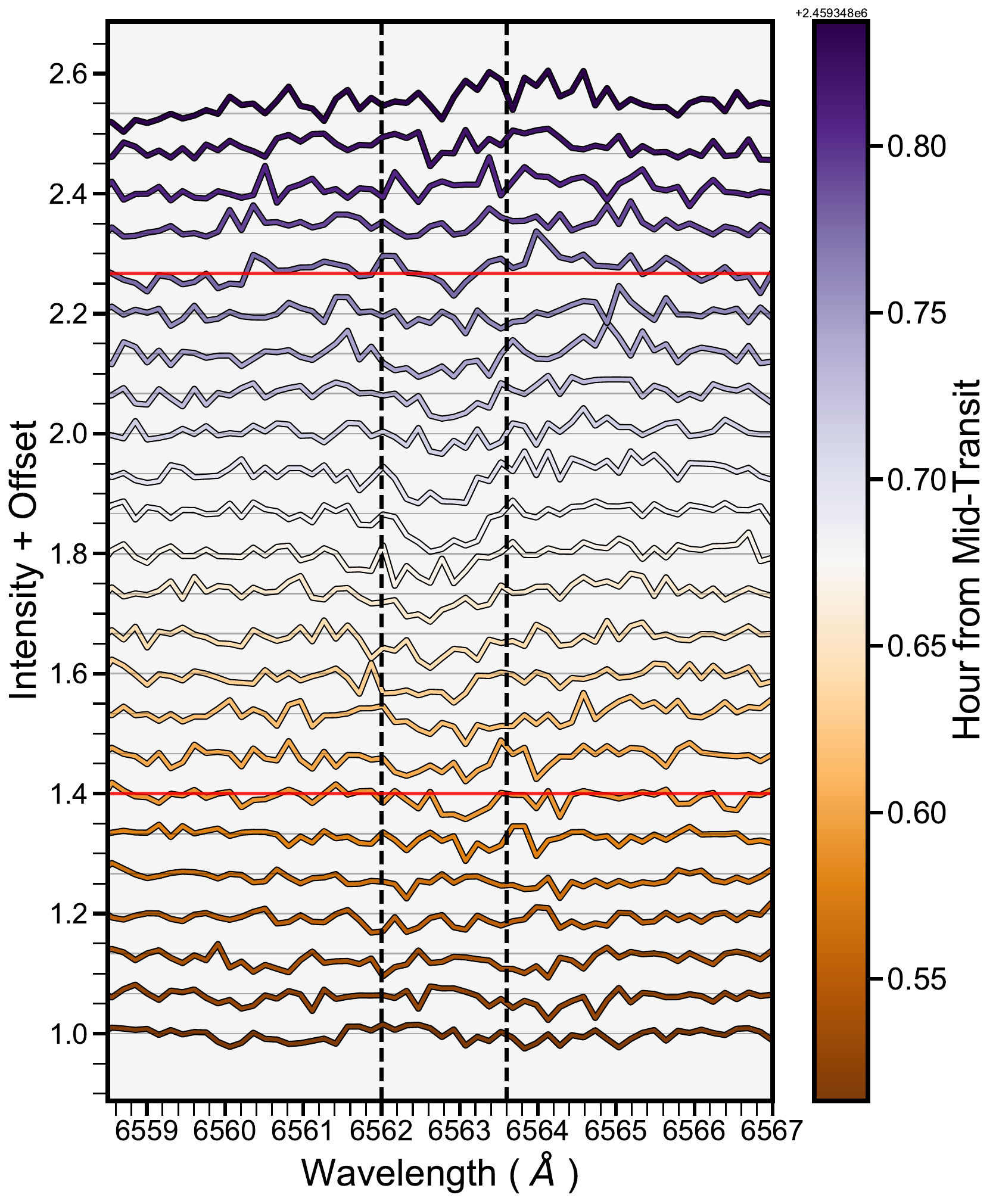} \caption{The time evolution of the reduced and template-corrected spectra for HIP\,67522\,b spanning observations on 2021-05-14 UT. The color bar shows the time of observation relative to the predicted midpoint of the transit. The lines in red denote ingress and egress. The vertical dashed lines mark the region used to calculate the equivalent widths.} \label{fig:hip_ts}
\end{figure}

\begin{figure}[ht]
\centering
\includegraphics[width=0.48\textwidth]{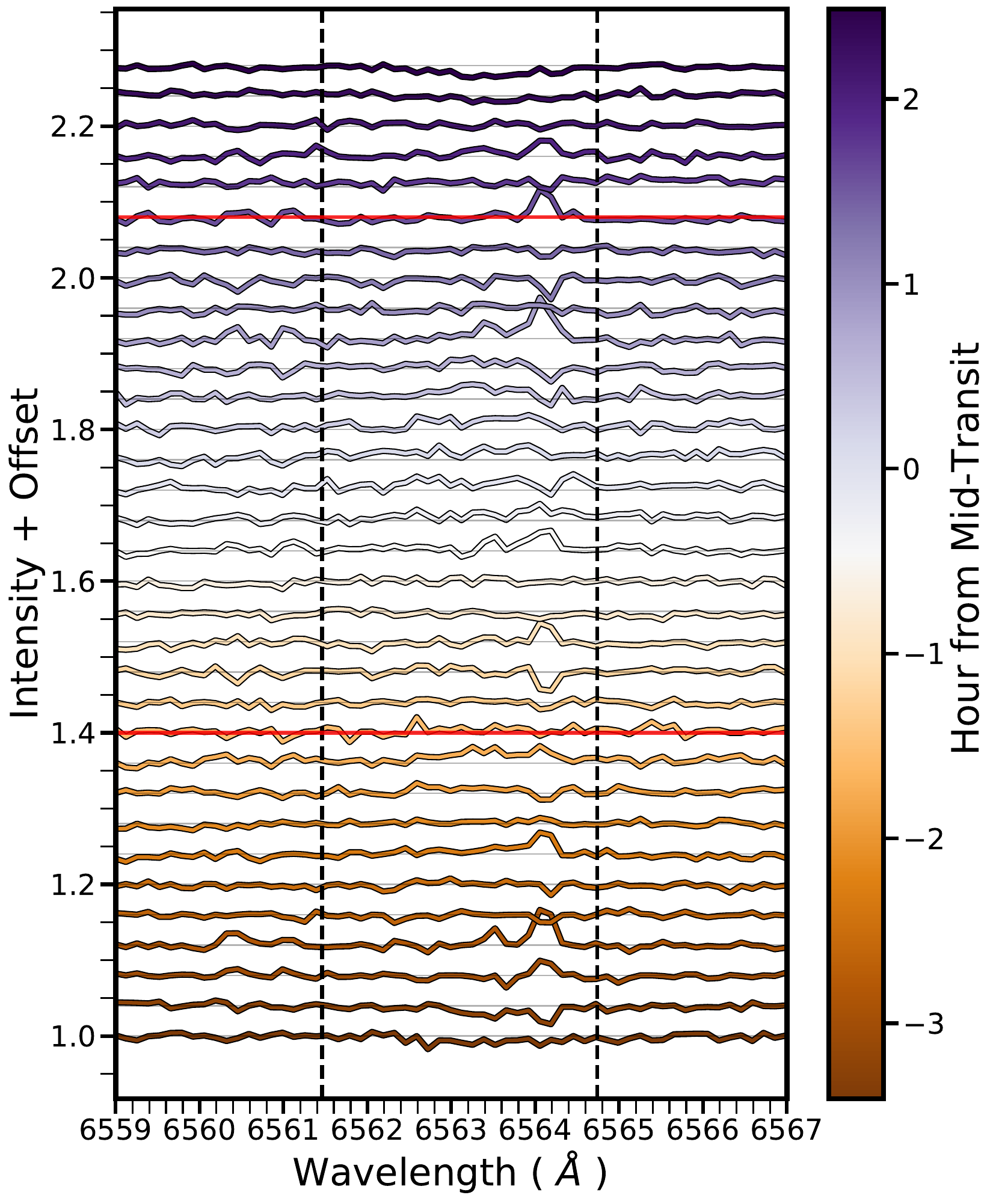} \caption{Same as Figure~\ref{fig:hip_ts} but for DS~Tuc~A\,b observations taken on 2019-10-07 UT.} \label{fig:ds_ts1}
\end{figure}

\begin{figure}[ht]
\centering
\includegraphics[width=0.48\textwidth]{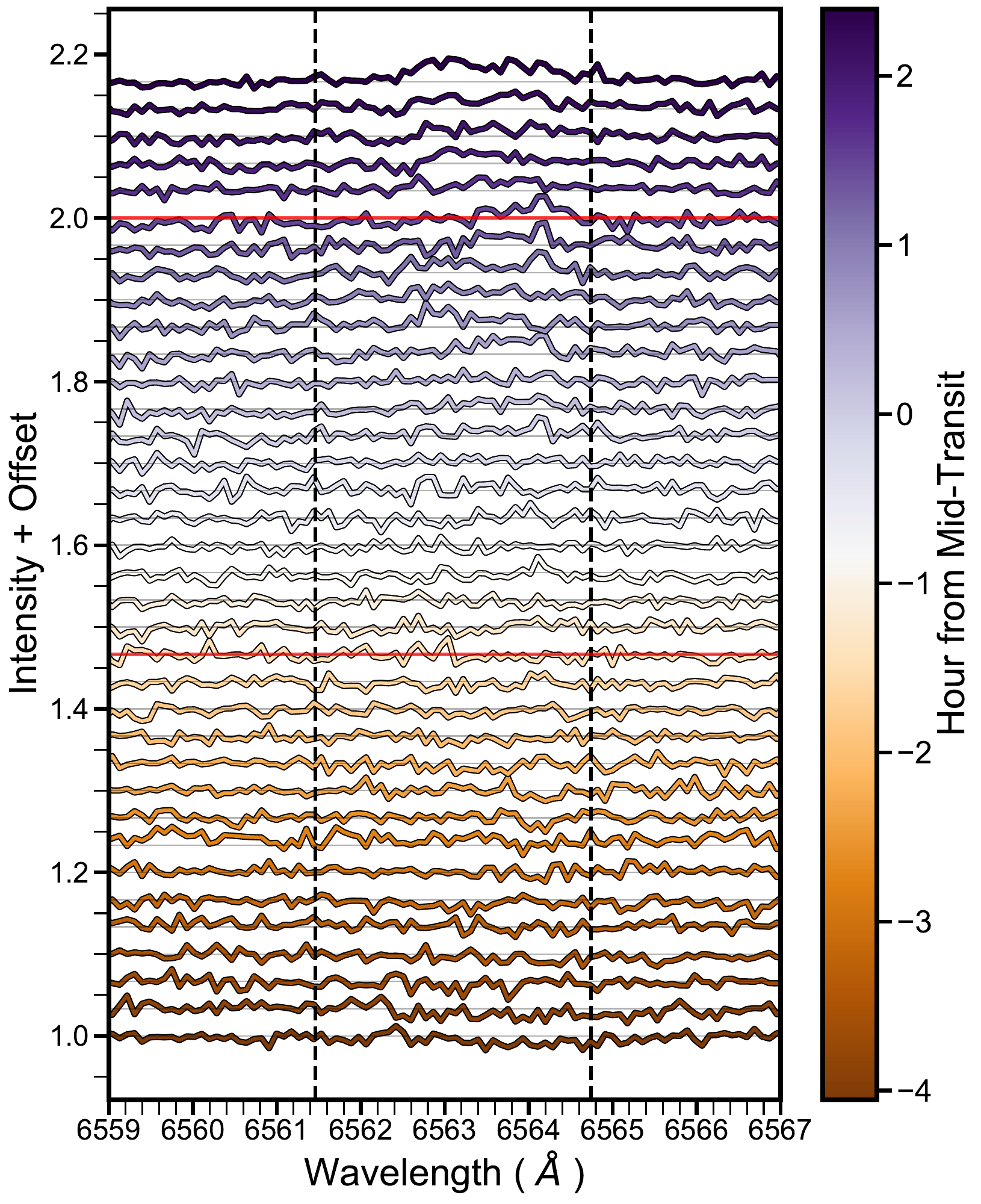} \caption{Same as Figure~\ref{fig:hip_ts} but for DS~Tuc~A\,b observations taken on 2019-08-19 UT.} \label{fig:ds_ts2}
\end{figure}

\begin{figure}[ht]
\centering
\includegraphics[width=0.48\textwidth]{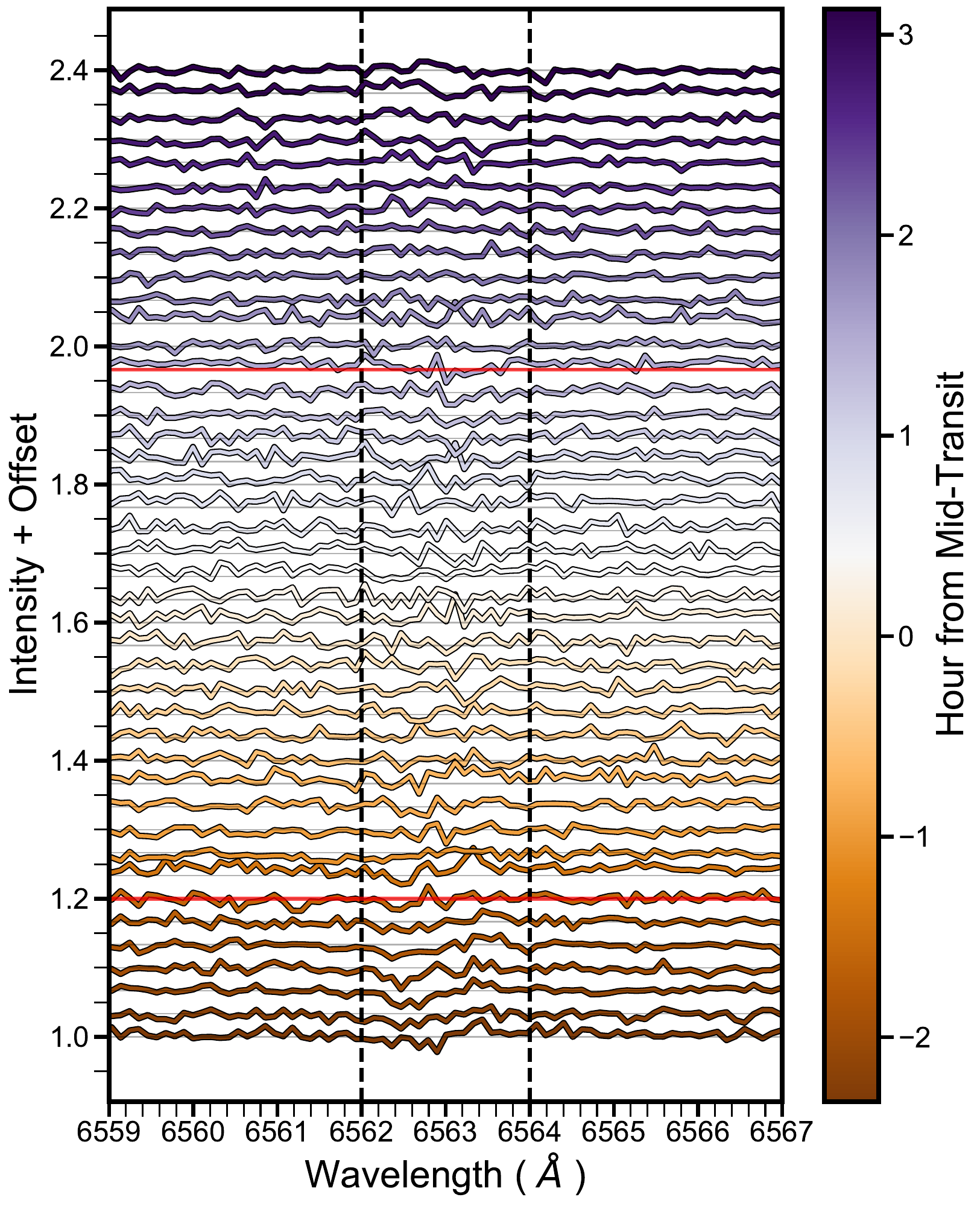} 
\caption{Same as Figure~\ref{fig:hip_ts} but for HD~63433\,b observations taken on 2020-03-07 UT.} \label{fig:hd_ts}
\end{figure}

We plotted the normalized and template removed time-series spectra and inspected the region near the center of the stellar H$\alpha$ feature. For each transit, we searched for significant changes in the spectra that coincided with the predicted times of transit. The time-series spectra  (after template removal) for each transit can be seen in Figures~\ref{fig:hip_ts}-\ref{fig:hd_ts}. Each show significant variations, primarily due to Poisson noise. Only HIP\,67522\,b shows a clear excess absorption (a feature below the flat-line prediction), at $\lambda\simeq6563$ and starting slightly after ingress and ending right at egress (Figure~\ref{fig:hip_ts}). For HD63433\,b and DS~Tuc~A\,b, no such features were seen (Figures \ref{fig:ds_ts1}, \ref{fig:ds_ts2}, and \ref{fig:hd_ts}).

\begin{figure}[t]
\centering
    \includegraphics[width=0.48\textwidth]{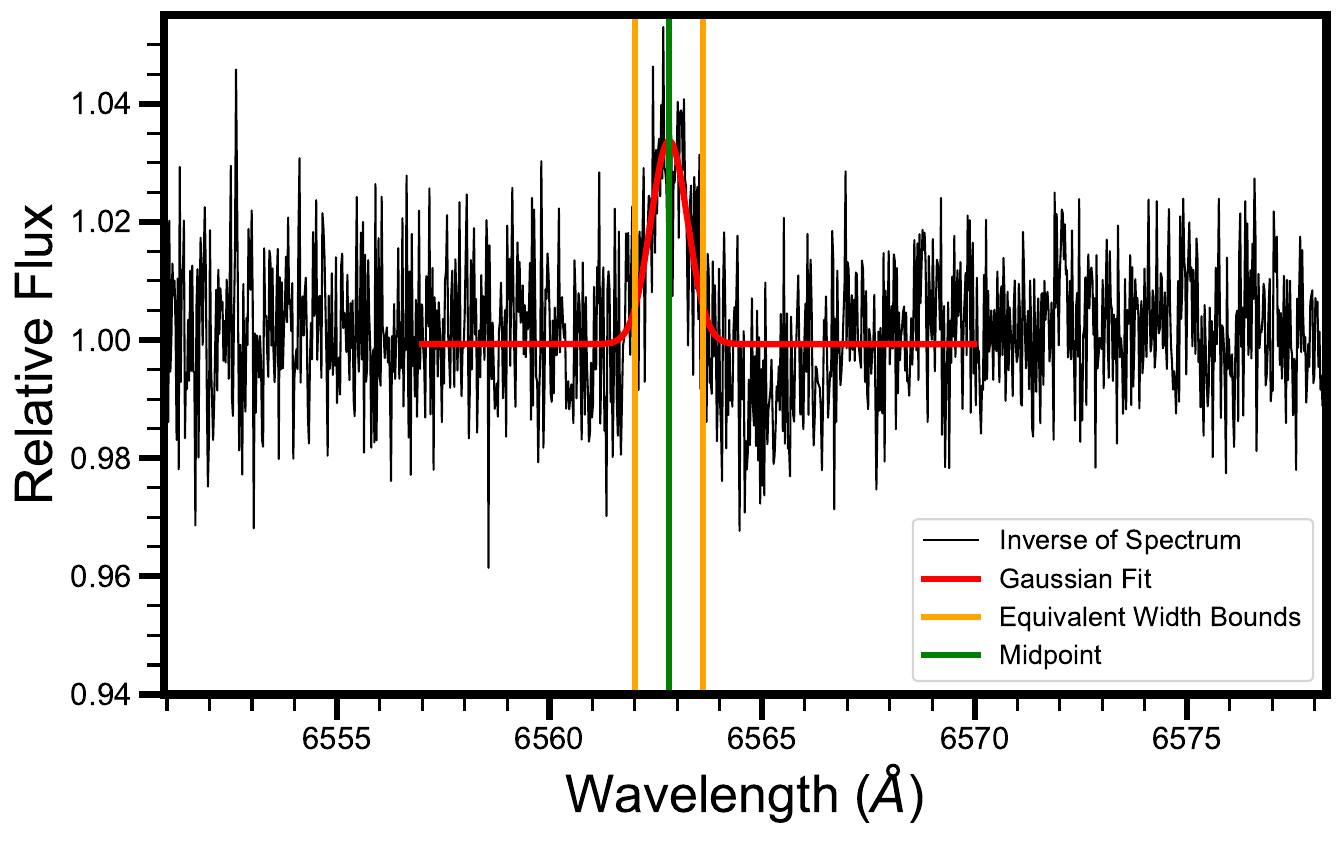} \caption{Median combination of the telluric and stellar corrected spectra of HIP67522 taken within the predicted transit times of HIP67522 b. We fit a Gaussian to the blueshifted absorption feature and use this feature to estimate the best region of integration for our equivalent width measurements.} \label{fig:hip_stack}
\end{figure}

As an additional test, we mean combined all of the in-transit data for each planet and transit. As above, we see a significant feature in HIP~67522\,b only. For this transit, we estimate the location and width of the line by fitting the stacked in-transit spectrum with a Gaussian (Figure~\ref{fig:hip_stack}), which we use to set the bounds for our equivalent width estimate.

Although some weak variation is seen for other systems, they are not consistent with planetary absorption. For example, DS Tuc A b shows excess emission, broad and centered on the stellar H$\alpha$ line (no blueshift). This is most likely due to stellar variability.

\subsection{Equivalent Width Measurement}\label{eq_width}

To calculate the equivalent widths, we used the python package \texttt{SpecUtils}. We provided \texttt{SpecUtils} with the spectral region of interest (Table~\ref{tab:obs}), and used the same spectral region for each time series spectrum of a given target. 

For HIP~67522\,b, we centered the equivalent width integration region on the narrow, blue-shifted absorption feature that appeared in-transit. This region was 1.6 \AA wide, spanning from 6562.0 \AA to 6563.6 \AA. 

Other targets had no visible in-transit excess absorption detected. The region for estimating the equivalent width was selected to encompass a majority of the H-alpha line based on an estimate of its full width at half maximum. This is approximately 1 - 3 \AA  in width, and varied based on the H-alpha profile for each particular target and instrument used for observation. The regions used are listed in Table~\ref{tab:obs}.

\subsection{Estimating Uncertainties}\label{sec:uncertainties}

Because all spectra are relatively high SNR, we expect the dominant source of uncertainty to come from the telluric removal, continuum normalization, and other systematics in the dataset. To estimate these uncertainties we employed an end-to-end Monte Carlo analysis, where we perturb the spectra and repeat our Telfit procedure. This method does not account for instrumental effects, which we assume are consistent across all spectra within a given observation. It also will not account for stellar variability, which we handle separately in our examination of the time-series. 

The first step was to adjust the spectra by the measurement uncertainties. Not all reduced data comes with estimated errors on the spectral intensity. For those, we estimated uncertainties on the intensity by using the square root of the flux counts per pixel. We also tested shifting the spectra by 1 pixel, subtracting, and calculating the standard deviation, which yielded similar but slightly higher uncertainties.

We perturbed each point along the spectrum assuming the noise is Gaussian. After perturbation, we repeated the full analysis where we continuum normalize, remove telluric features, remove stellar features, and measure the equivalent width (Section~\ref{sec:methods}). This process was repeated 100 times for each equivalent width measurement of each spectrum. We adopted the standard deviation of these 100 estimates as our uncertainty estimate. 

\subsection{Comparing with Calcium}\label{sec:calcium}

As a check of our approach and uncertainties, we carry out an identical analysis on the Ca line centered at 6102.8\AA. HIP\,67522\,b is known to have strong spectral features, but the Ca feature here is expected to be well below detection limits even for this planet \citep{Thao2024b}. The feature should be even weaker in the other two planets, as they have higher surface gravities. This region is less sensitive to stellar variations than H$\alpha$, so this comparison mostly tells us about systematics not related to the host star (e.g., from the instrument or our procedure).

 We generate a stellar template of this region as we did with our H$\alpha$ analysis. Continuum normalization and telluric correction follow similarly. There are fewer telluric lines present in the region, so \texttt{TelFit} parameters are more poorly constrained, but achieves a reasonable fit to the data.

\begin{figure}[ht]
    \centering
    \includegraphics[width=0.48\textwidth]{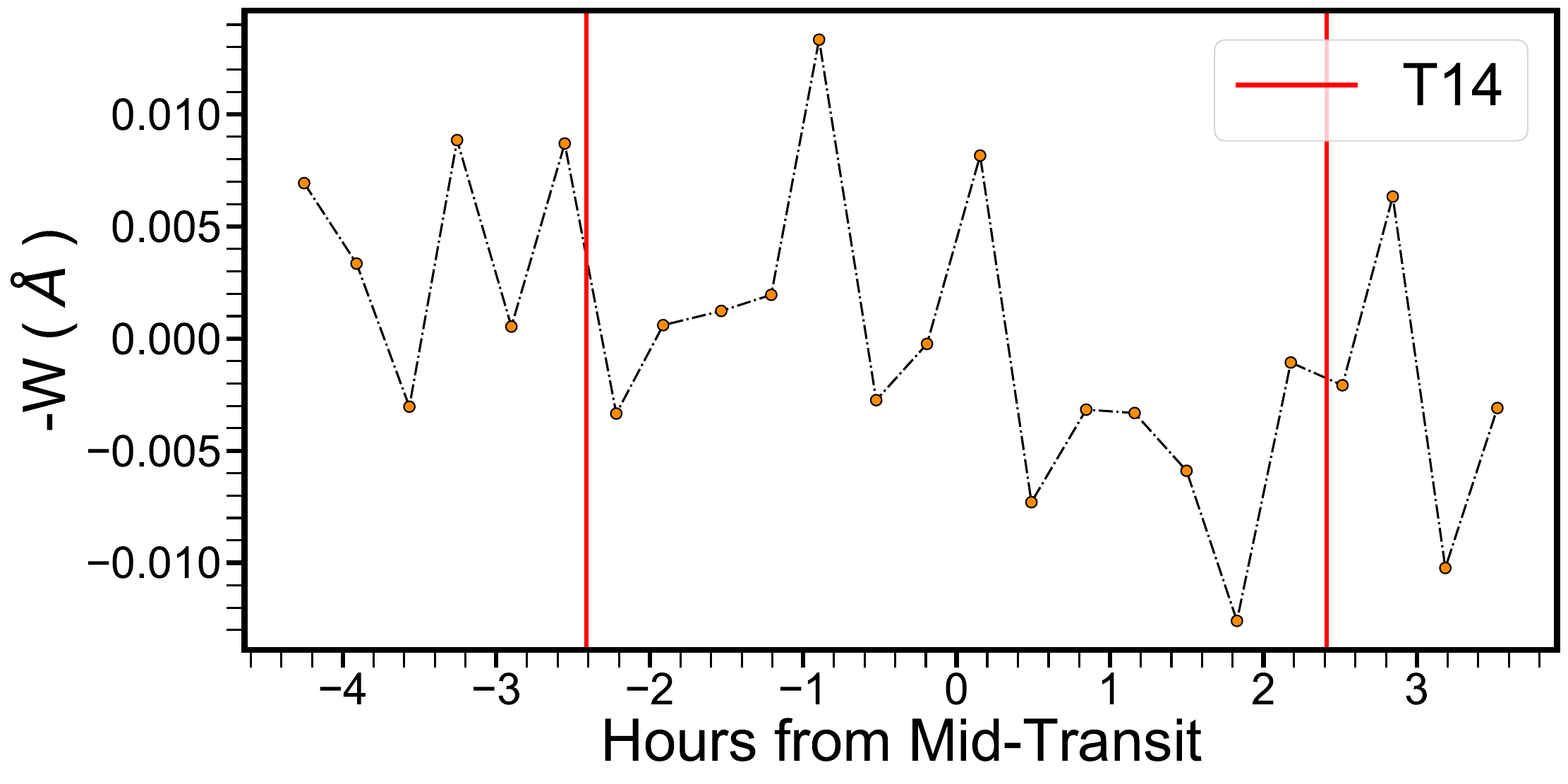} 
    \includegraphics[width=0.48\textwidth]{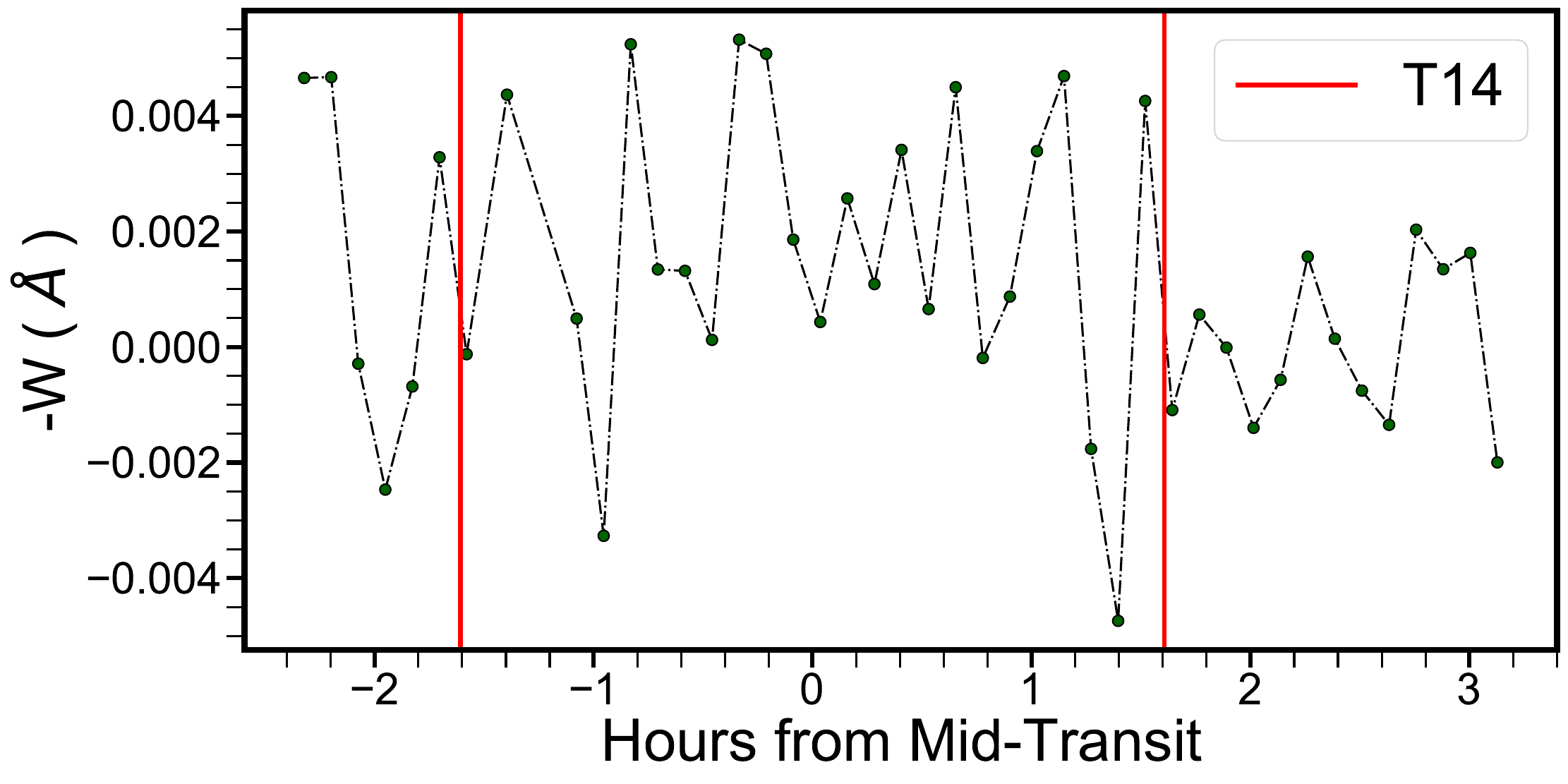}
        \includegraphics[width=0.48\textwidth]{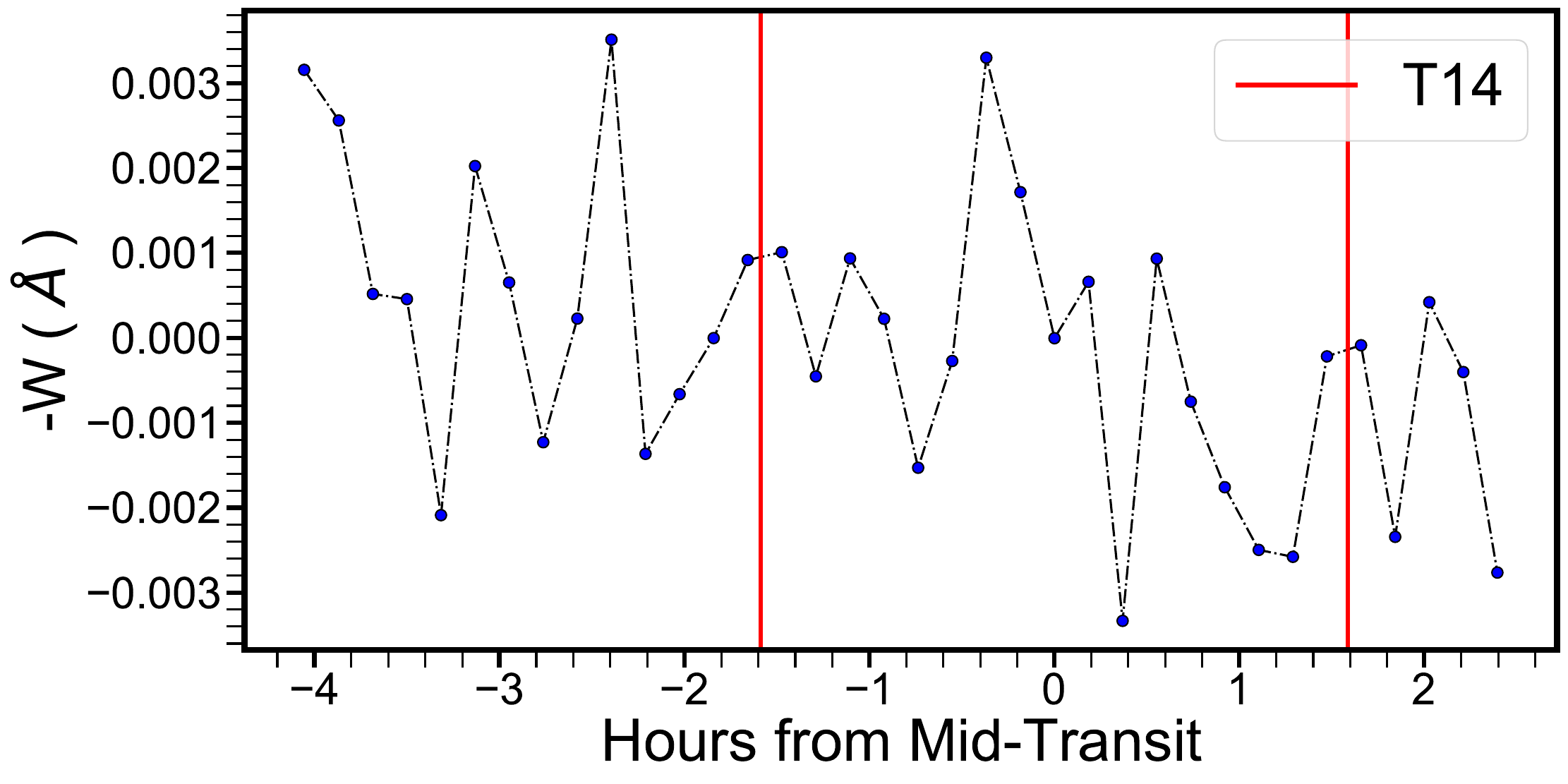} 
    \includegraphics[width=0.48\textwidth]{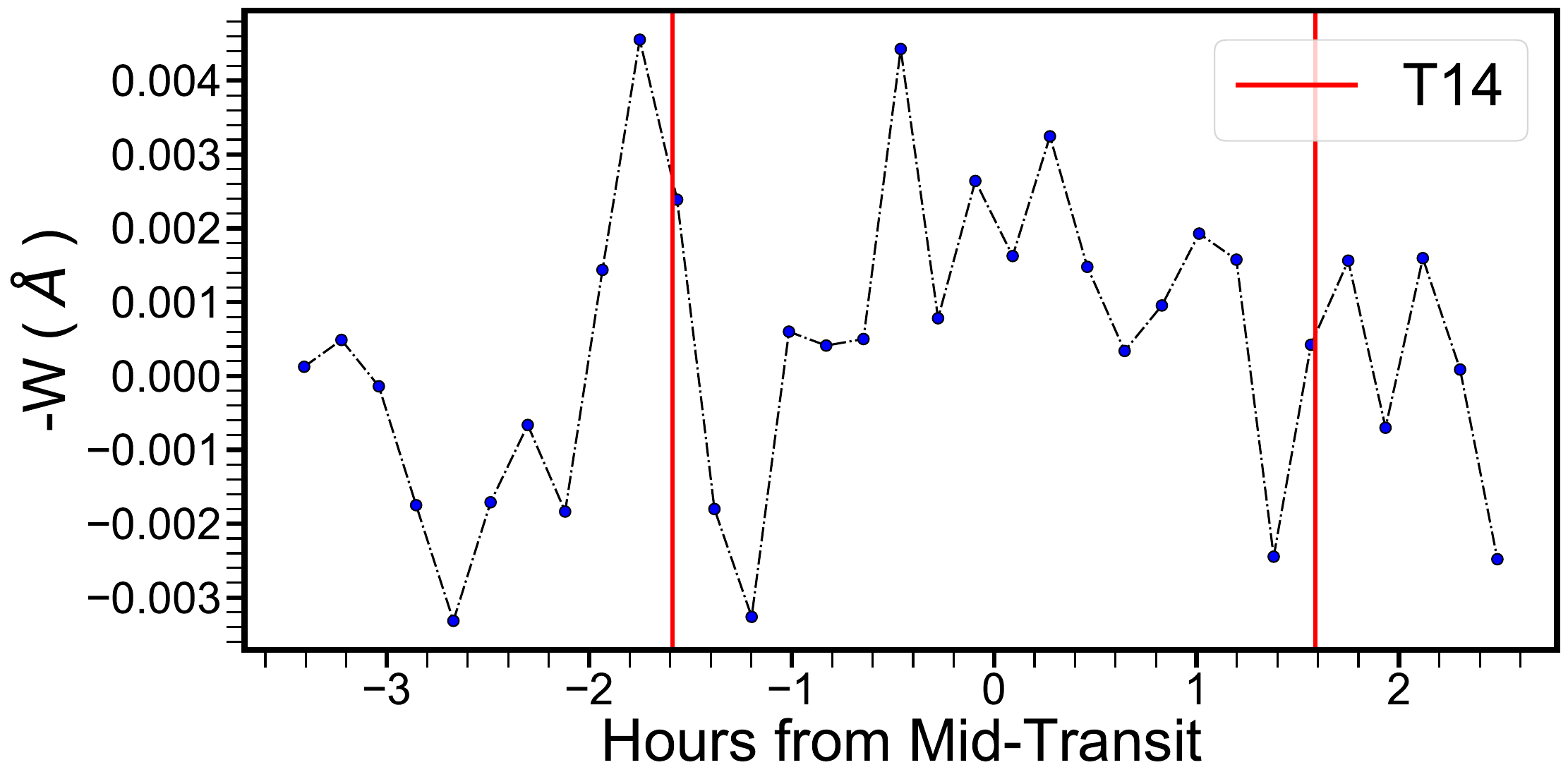} 
\caption{The equivalent widths of the Ca I absorption feature located near 6103 \AA plotted against the time of observation in relation to the predicted transit times for HIP67522 b, HD63433 b, and two nights for DS Tuc A b in descending order. The transit ingress and egress are marked by the vertical red lines. This feature is relatively stable within uncertainties, indicating no detections of atmospheric calcium.   } \label{fig:calcium}
\end{figure}

 When we compare the equivalent width measurements against the times of observations and the predicted transit ingress and egress, we find the Ca feature is stable throughout the course of observation Figure~\ref{fig:calcium}. Further, the point-to-point variation is comparable to, or less than the estimated uncertainties, suggesting our assigned uncertainties are reasonable or slightly conservative.

\section{Results}\label{sec:results}

\subsection{HIP\,67522\,b}

The HIP\,67522\,b detection yielded an excess absorption (depth) of 3.44$\pm$0.28\%, a SNR of 12. HIP\,67522\,b was also the only target of the three that showed an excess H$\alpha$ detection consistent with coming from the planet. This is most clear in Figure~\ref{fig:hip_stack}, which shows the (inverted) template- and telluric-removed in-transit spectra. There is a clear Gaussian excess at $\simeq6562.8\AA$. 

\begin{figure}[htb]
\centering
\includegraphics[width=0.48\textwidth]{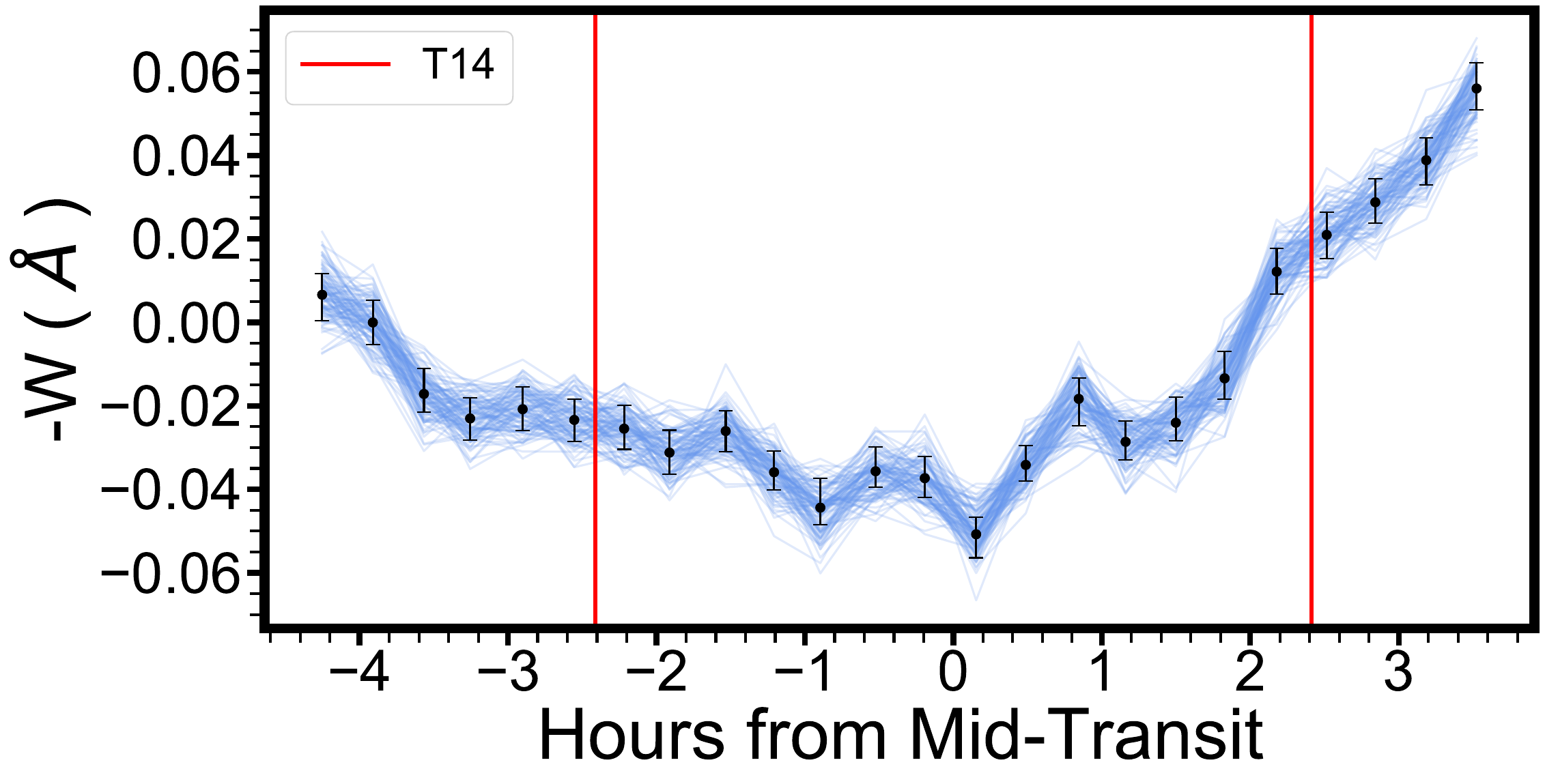} \caption{Equivalent widths of H$\alpha$ for HIP\,67522\,b spanning observations on 2021-05-14 UT. Each of the 100 overlapping blue lines represents a single set of realizations from our Monte-Carlo error perturbation. The black dots are the average equivalent width for each time step, with error bars marking a 1$\sigma$ range of uncertainties. The vertical red lines denote the start and end times of the predicted transit (T14).} \label{fig:hip_eqws}
\end{figure}

Like most young stars, HIP\,67522 shows significant H$\alpha$ variability. This can be seen in the equivalent width versus time (Figure~\ref{fig:hip_eqws}) as significant out-of-transit variation as well as a significant offset between pre-transit and post-transit H$\alpha$ levels. The difference in pre- and post-transit H$\alpha$ equivalent widths is comparable to the in-transit dip. This effect is not seen in the Ca analysis (Figure~\ref{fig:calcium}).

The results for this system strongly suggest the detection is associated with the planet and not the host star: 1) the excess is blue-shifted with respect to the stellar line by $21\pm3$\,\kms, 2) the line width is narrow ($<$15\,\kms) compared to the stellar line (51\kms), and 3) the line profile does not shift in position during the transit. 

For (1), $21\pm3$\kms{} is large compared to global circulation models for hot Jupiters, which generally predict shifts of 3-5\kms{} or up to 10\kms{} in the most extreme cases \citep{Showman2013,Tan2019}. However, these are not applicable when the material becomes unbound, after which stellar winds become more important than those on the planet. For escaping material, hydrodynamic or Parker-type wind models would be more appropriate, and these predict velocities for the upper atmospheres and outflow velocities of a 5-10\kms{} at escape altitudes, with larger velocities possible in the presence of stellar winds \citep{Murray-Clay2009, VillarrealDAngelo2018}. Indeed, models of He escape in young planets \citep{Allan2025} predict velocities of 20-60\kms, and a wide range of prior observations of escaping material have found velocities of 20-100\kms{} \citep[e.g.,][]{Murray-Clay2009, Ehrenreich2015}. Thus, $21\pm3$\,\kms{} is on the high end of expectations for planetary winds, but well within predictions for escaping material, especially given HIP 67522's young age and it's expected stellar wind. 

The high blueshift also disfavors a stellar origin. Reproducing the blueshift signal would require an surface feature (spot, faculae, or flare) at $\sim$40\% of the projected stellar radius, which would traverse $\gtrsim$20\% of its longitudinal path during our observations (the rotation period is just 1.4\,days). This would yield a velocity shift of at least 20\kms{} over the observing window. While suspended material (prominences) might yield a more stationary feature, these manifest as emission rather than absorption \citep[e.g.,][]{Donati1999,Waugh2021}. Flares and coronal-mass ejections also tend to create non-Gaussian and asymmetric profiles in time and wavelength \citep{Howard2023}.

One piece of evidence disfavoring a planetary detection is that the event duration is shorter than expected, with a significantly late ingress and marginally early egress. This can be seen in Figure~\ref{fig:hip_eqws}. 

As an additional test we analyzed another dataset for HIP\,67522\,b, taken with the same spectrograph and in a similar manner on 2023-02-28 UT. The advantage of this dataset is that it was accidentally taken one night off from the transit, meaning it was observed as though there might be a transit, but none is present. Any detection therefore would suggest an activity origin. 

\begin{figure}[ht]
\centering
\includegraphics[width=0.48\textwidth]{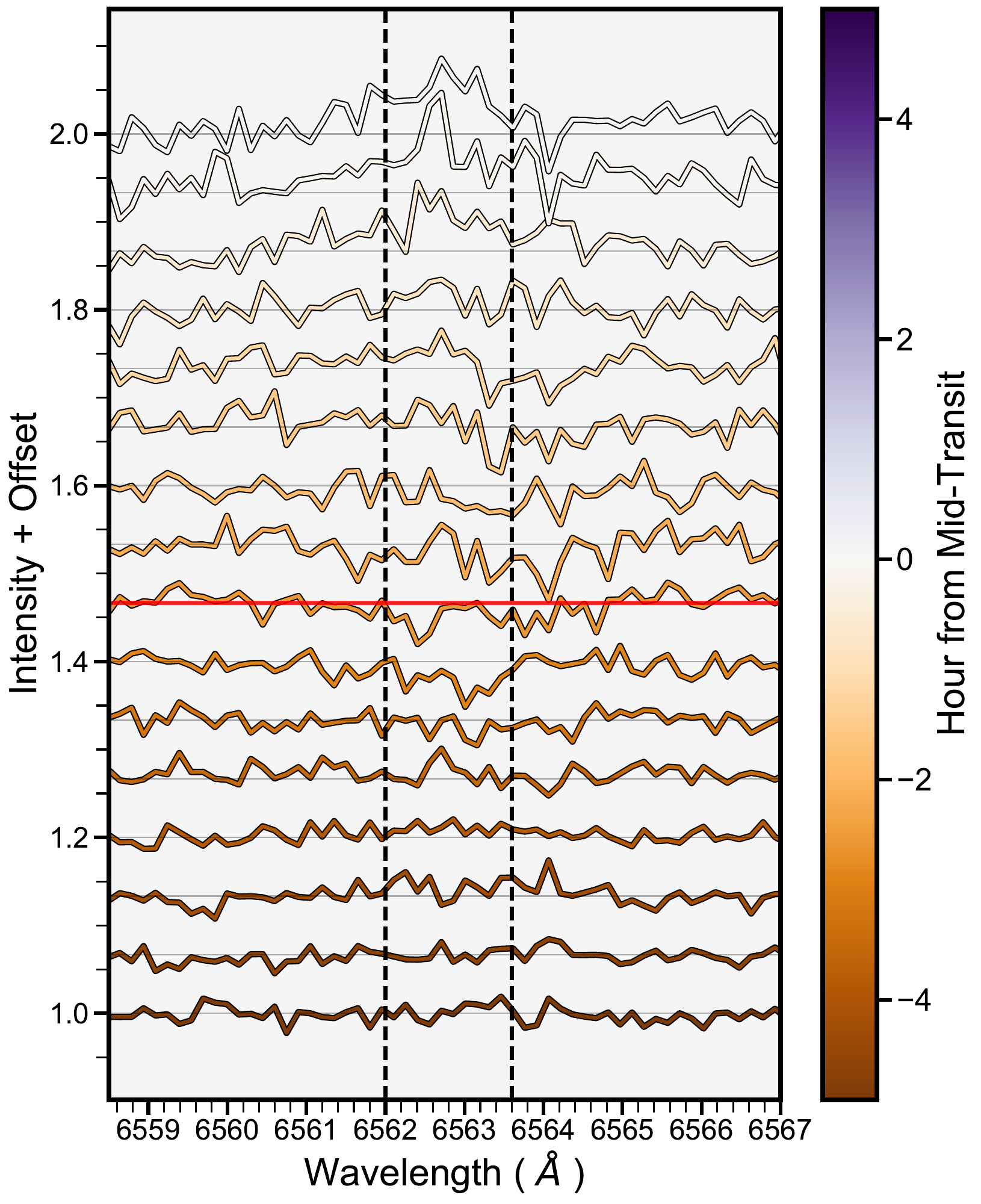} \caption{The time evolution of the reduced and template-corrected spectra for HIP\,67522\ spanning observations on 2023-02-28 UT. The observations occurred 1 day after the predicted transit of HIP\,67522\,b, resulting in a planet-less series of observations. The color bar shows the time of observation relative to the simulated midpoint of the transit had observations spanned the actual predicted transit time. The line in red denotes ingress of the actual transit time +1 day. The vertical dashed lines mark the region used for the in-transit feature.} \label{fig:hip_other}
\end{figure}

Our analysis was identical to the other data (just adding 1 day to the transit time). The resulting time-series template-subtracted spectra can be seen in Figure~\ref{fig:hip_other}. The transit-free time-series spectra {\it do not} show an excess absorption feature like what we see in the transit observations. Compared to the true transit observations, the  resulting feature is in emission, asymmetric (non-Gaussian), and much broader in wavelength. This more closely resembles that of HD 63433 and DS Tuc than the HIP\,67522\,b transit. However, the intrinsic scatter of H$\alpha$, even for a narrow wavelength range, is comparable to the transit detection. This mostly re-affirms the initial detection, albeit not conclusively. 

\subsection{HD~63433\,b \& DS~Tuc~A\,b}

\begin{figure}[t]
\includegraphics[width=0.48\textwidth]{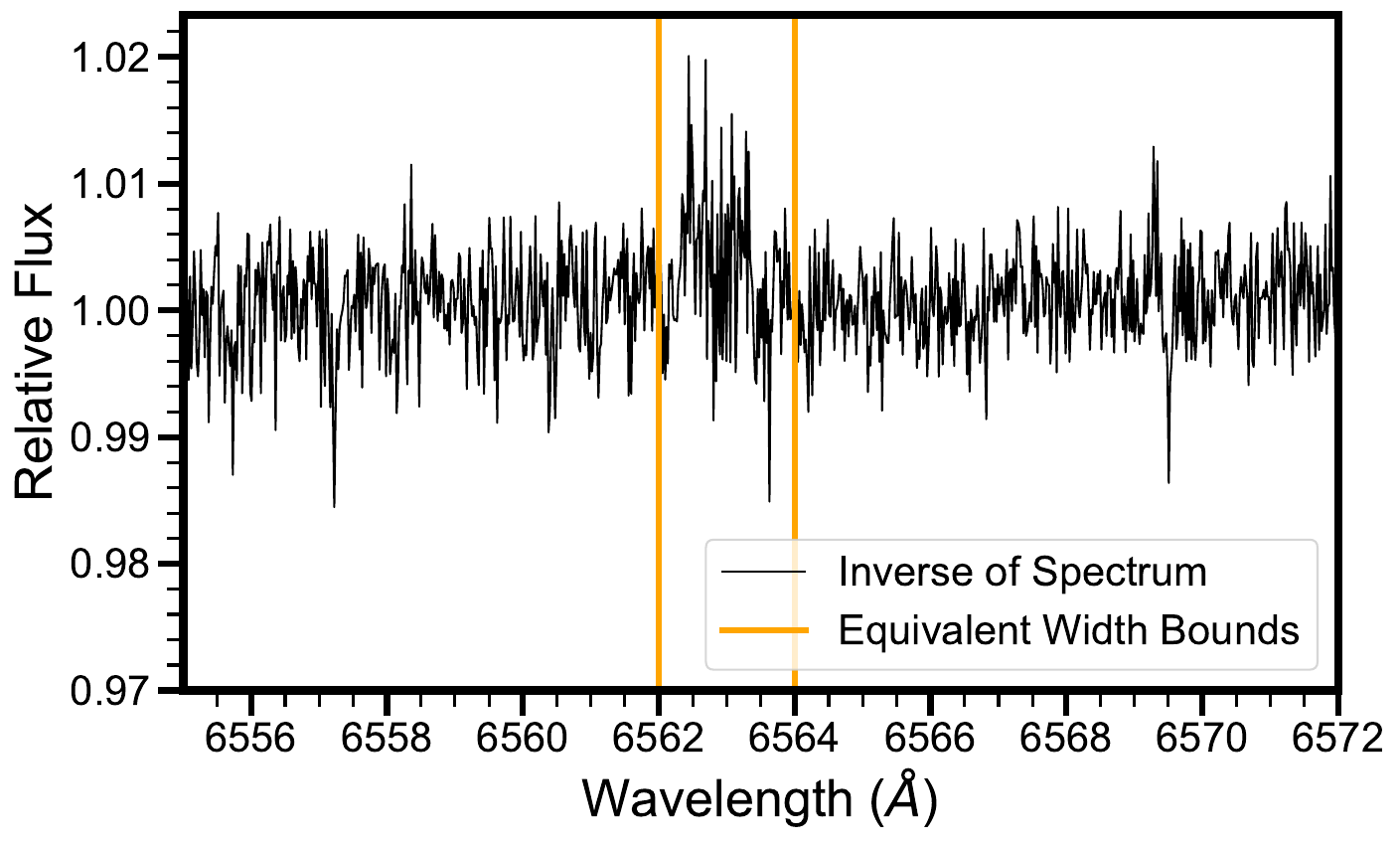} 
\caption{Same as Figure~\ref{fig:hip_stack}, but for HD\,63433\,b. A feature consistent with excess absorption is visible, but the profile and location are consistent with a stellar origin. } \label{fig:hd_stack}
\end{figure}

\begin{figure}[t]
\includegraphics[width=0.48\textwidth]{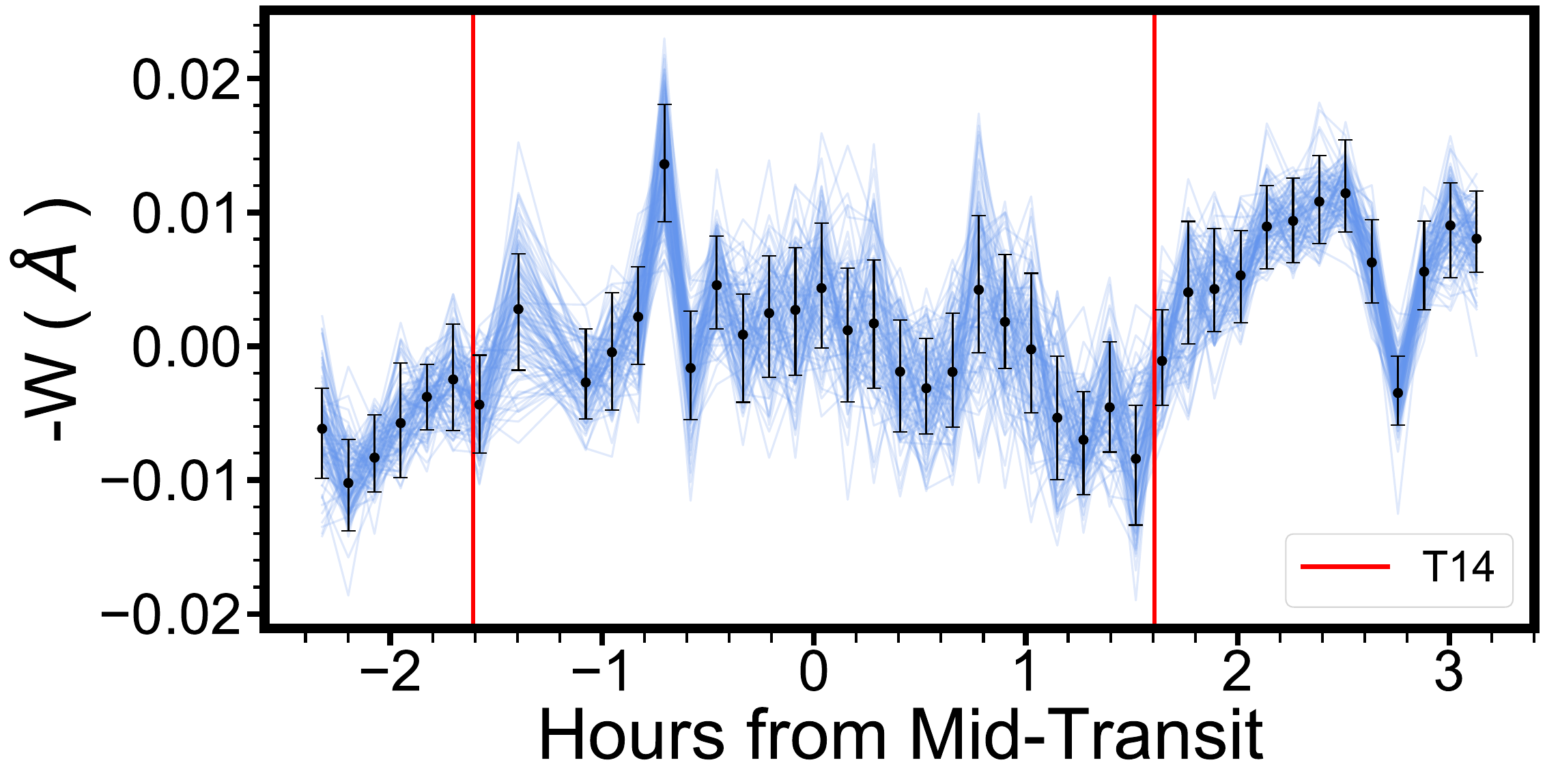} \caption{Same as Figure~\ref{fig:hip_eqws} for HD\,63433\,b. A marginally significant dip is seen near transit egress, but this is unlikely to be associated with the planet given the timing and resulting stacked H$\alpha$ profile (Figure~\ref{fig:hd_stack}).} \label{fig:hd_eqws}
\end{figure}

For HD~63433\,b, fitting the excess absorption gave a depth of 0.85$\pm$0.20\%. This suggests a significant (SNR$\simeq$4) detection, but we attribute this to the star rather than the planet. The stacked transit spectra show a non-Gaussian profile, sits close to the host H$\alpha$ line, and is as broad as the stellar lines (Figure~\ref{fig:hd_stack}). The signal is also entirely driven by a small increase in the absorption for the last $\simeq$30\,minutes of a nearly 4-hour transit (Figure~\ref{fig:hd_eqws}). Variations of similar amplitude are seen in other parts of the time-series, and the time-resolved spectra (Figure~\ref{fig:hd_ts}). A forced-fit to a narrower planet-like signal as seen for HIP~67522\,b suggested a limit on the depth of HD~63433\,b H$\alpha$ absorption of $<0.9$\% at 3$\sigma$. 

\begin{figure}[t]
\includegraphics[width=0.48\textwidth]{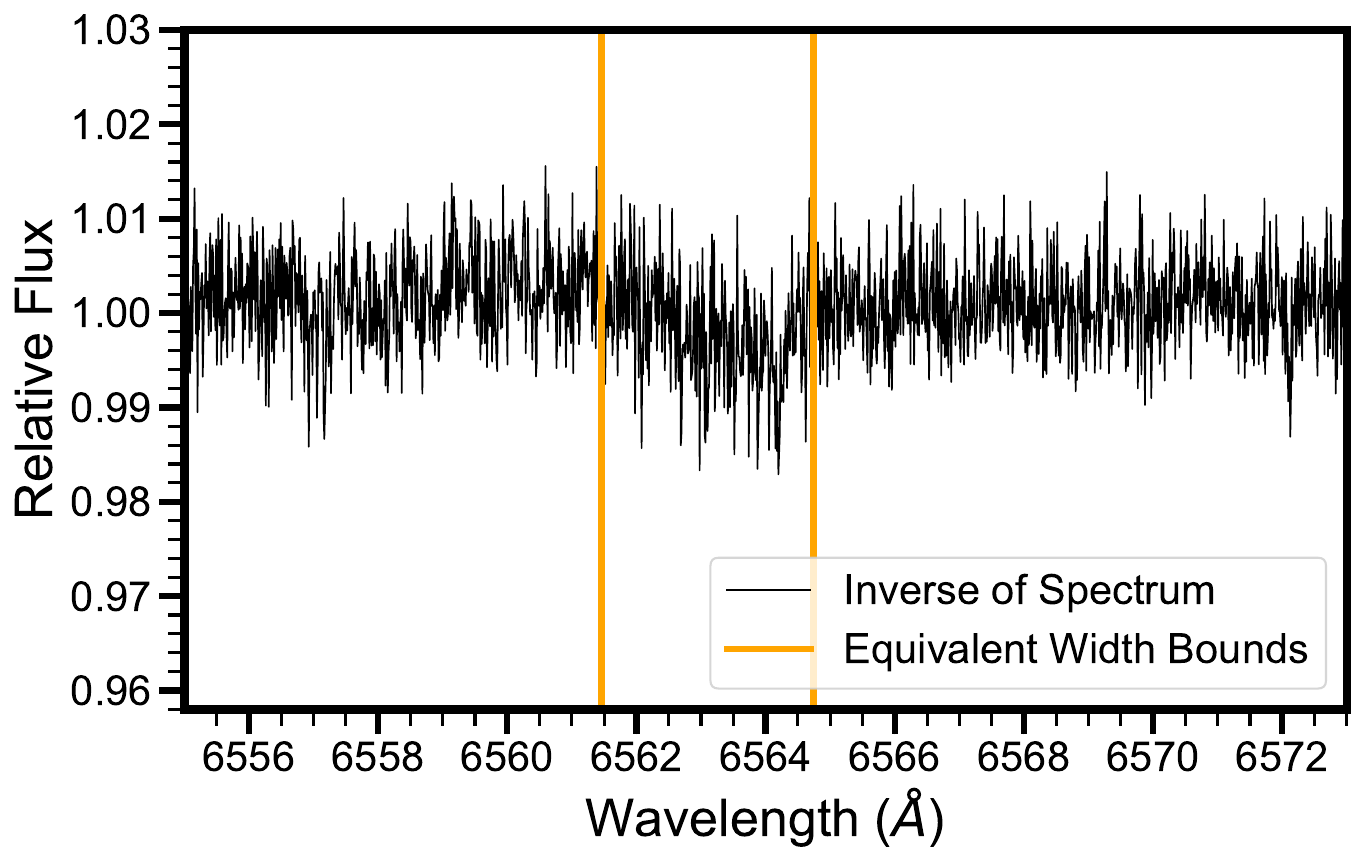} 
\includegraphics[width=0.48\textwidth]{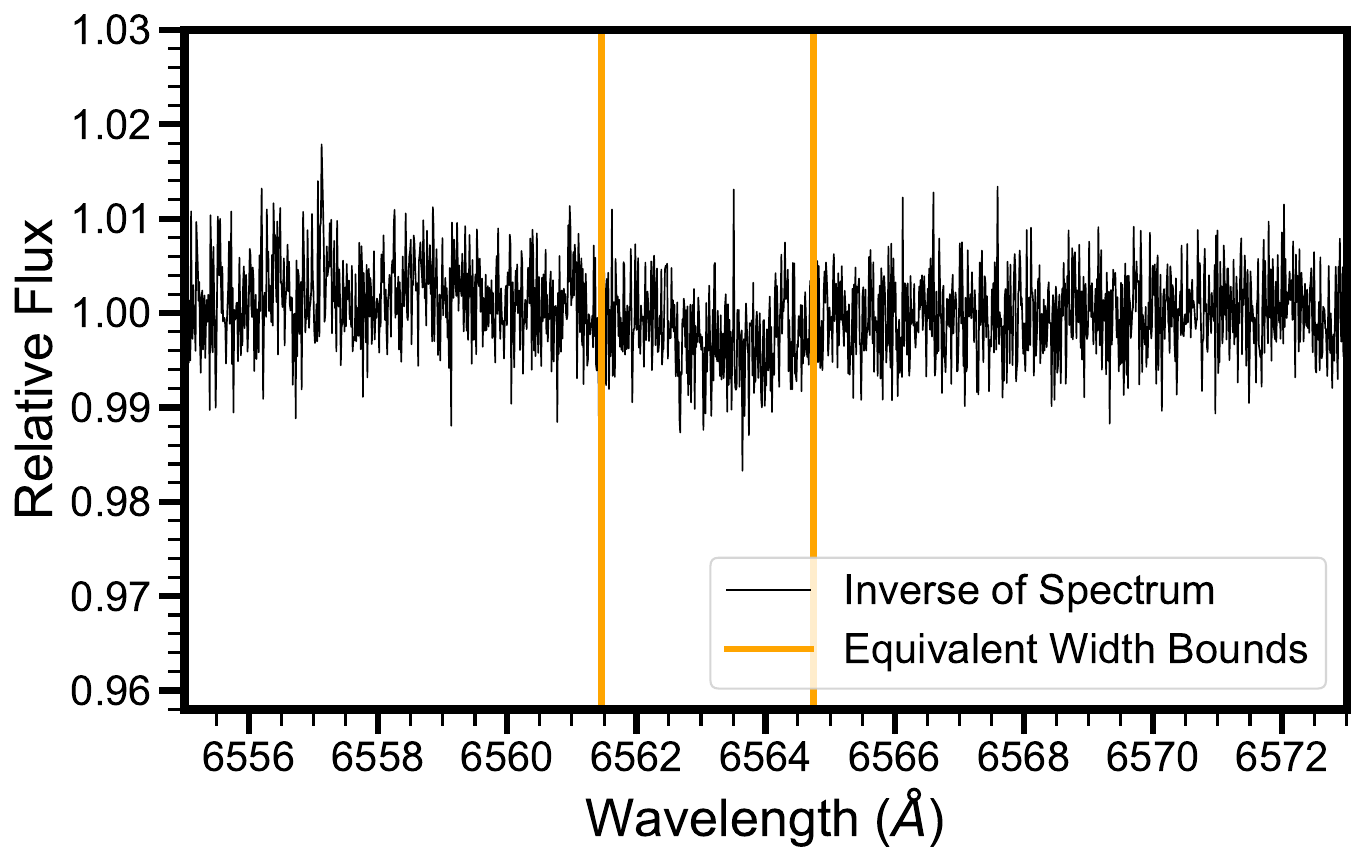} 
\caption{Same as Figure~\ref{fig:hip_stack} but for DS\,Tuc\,A\,b for the two different transits observed. Top is 2019-08-19 and bottom is 2019-10-07 UT. The orange lines indicate the bounds of integration for the equivalent width calculation (Figure~\ref{fig:ds_eqws}). A feature is present, but it shows excess emission (weaker absorption) - an exosphere would manifest as excess absorption.} \label{fig:DS_stack}
\end{figure}

\begin{figure}[t]
\includegraphics[width=0.48\textwidth]{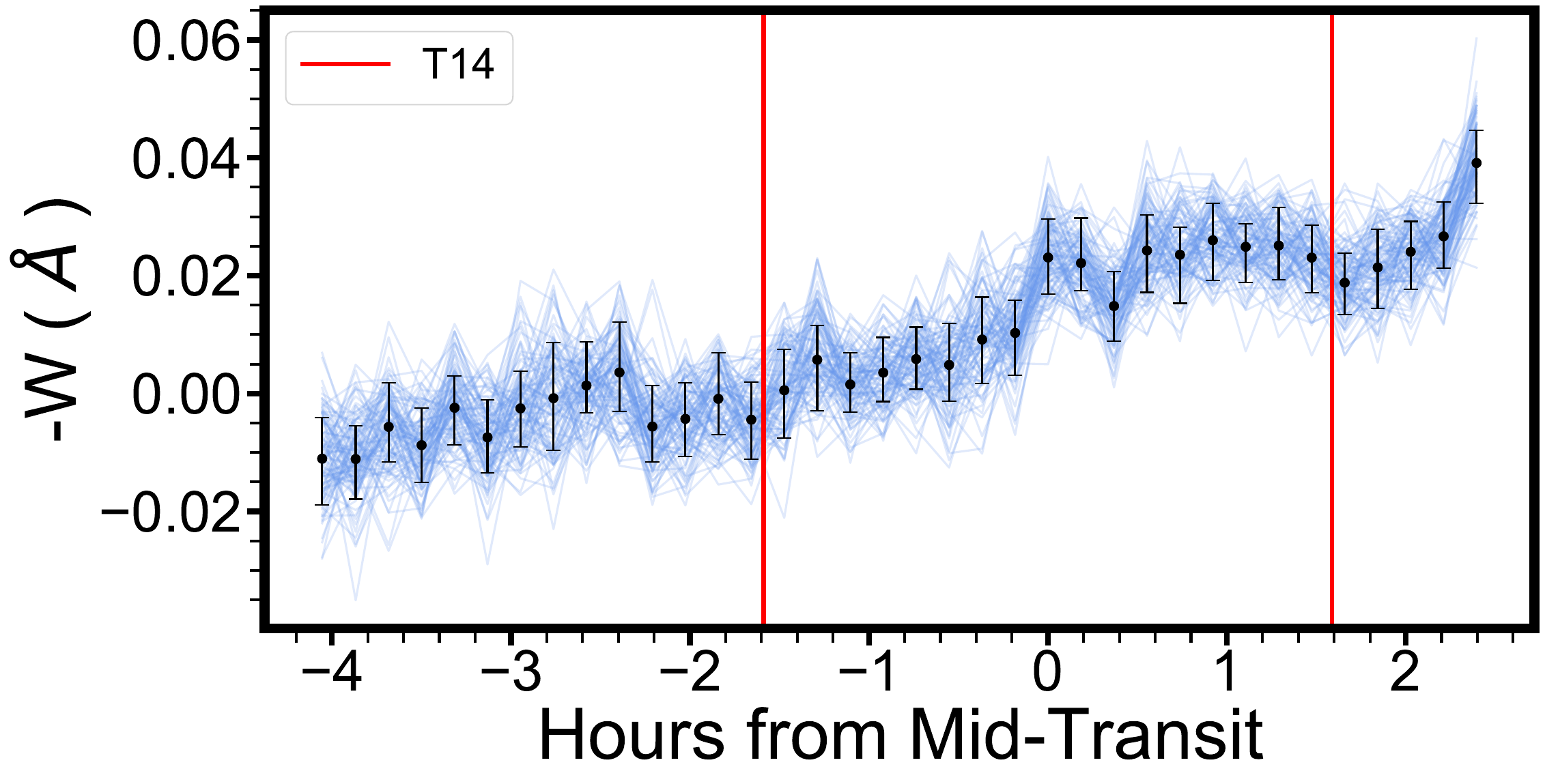} 
\includegraphics[width=0.48\textwidth]{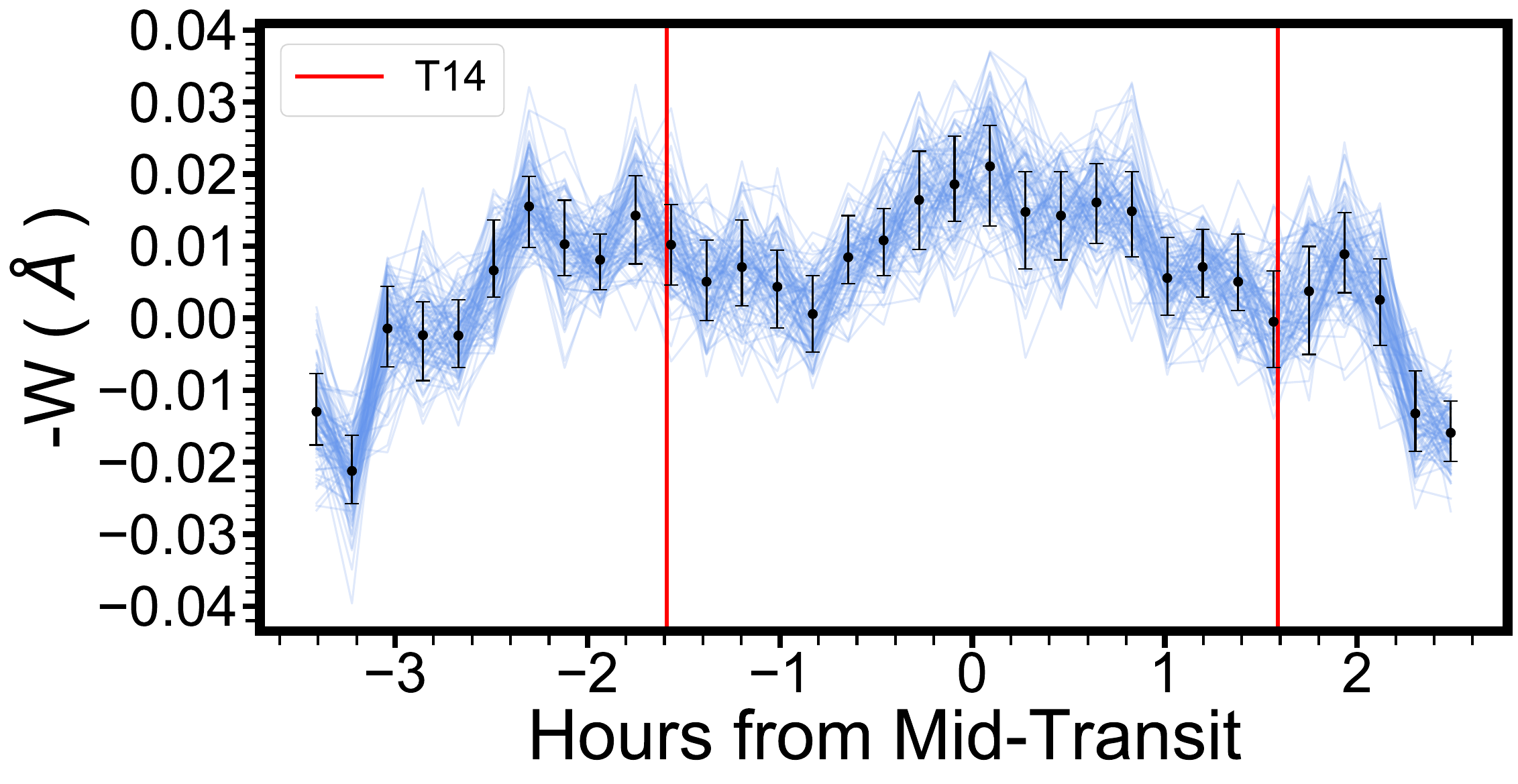} 
\caption{Same as Figure~\ref{fig:hip_eqws} for the two transits of DS\,Tuc\,A\,b. The top is 2019-08-19 UT and the bottom is 2019-10-07 UT.} \label{fig:ds_eqws}
\end{figure}

For DS~Tuc~A\,b, the detection shows excess {\it emission}, instead of absorption (Figure~\ref{fig:DS_stack}) - the opposite of what we expect for a transiting exosphere. Like HD\,63433\,b, the stacked H$\alpha$ is non-Gaussian, asymmetric, broad, and centered close to the stellar H$\alpha$ line. Also, the equivalent widths show similar variation in and out of transit (Figure~\ref{fig:ds_eqws}). The results strongly suggest the observed signal is stellar in origin. Accounting for this suggests an upper limit on the H$\alpha$ depth from the planet of 0.3\% at 3$\sigma$.

\section{Summary \& Conclusions}\label{sec:conclusions}

We analyzed time-series optical spectra of four young systems, HIP\,67522, HD\,63433, and DS\,Tuc\,A, all taken during transit as part of efforts to measure the spin-orbit alignment of the planetary system. Our goal was to look for excess absorption in H$\alpha$, a signature of ongoing atmospheric escape, which may be strong in such young systems. 

After careful removal of both telluric lines and the stellar spectrum, we find evidence for ongoing mass-loss in the youngest planet HIP\,67522\,b (17\,Myr). The excess absorption feature narrower and sits at a bluer wavelength than the stellar line. Our comparison transit-free time-series spectra also yield no feature consistent with our true-transit observations. These affirm that the signal is likely from the planet.

The timing of the transit feature, however, is off; the H$\alpha$ transit starts $\lesssim1$\,hour after the predicted ingress and ends $\lesssim$30\,minutes before egress (Figure~\ref{fig:hip_eqws}). The amplitude of the transit is similar to variations in the out-of-transit equivalent widths. The transit-free spectra also exhibit a large intrinsic variability in the equivalent widths. Together, the data favor a planetary origin, but we cannot completely rule out micro-flaring and other sources of variability.  Additional transits will be needed to check and further validate that the signal is from the planet.

We report no evidence of evaporative mass-loss for HD\,63433\,b and DS\,Tuc\,A\,b from analyzing H$\alpha$ absorption during transit. While both show weak features during the transit, none of them meets expectations for a planetary signal. DS\,Tuc\,A\,b has two transits, both of which are non-detections. It is possible another tracer (Ly$\alpha$, He) may provide evidence for atmospheric escape in these systems.

\begin{figure}[t]
\includegraphics[width=0.47\textwidth]{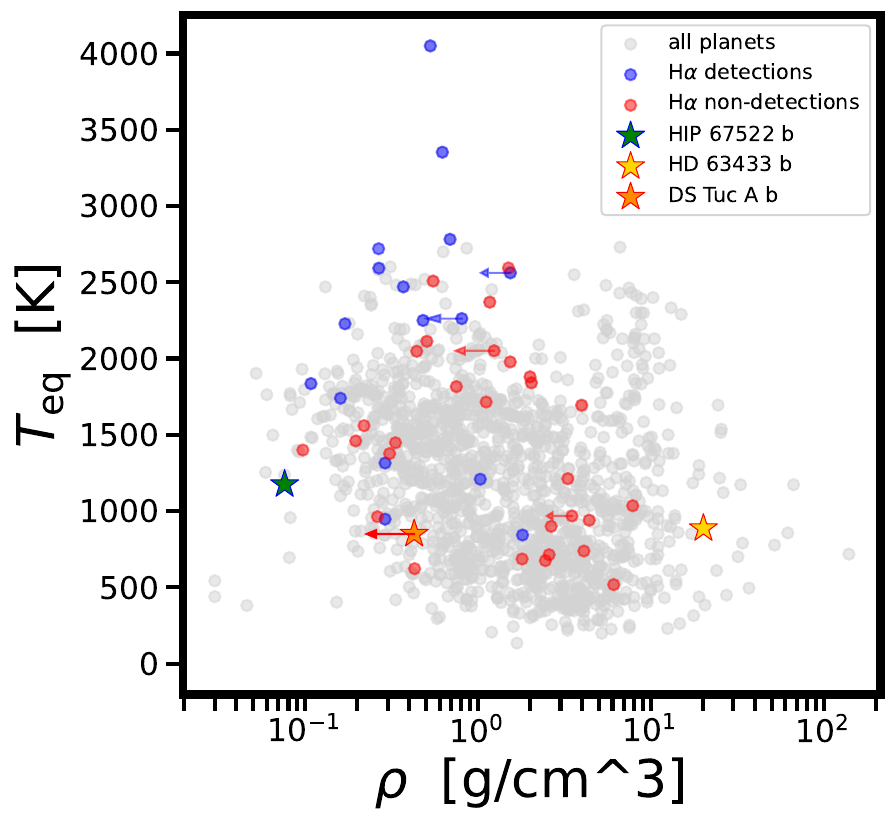} \caption{Planetary equilibrium temperature versus planetary density for targets with H$\alpha$ observations. Detections are shown in blue, with non-detections in red; the three targets considered here are shown as colored stars (outline for detection/non-detection). This plot suggests the major deciding factor is low $\rho$ and high $T_{\rm{eq}}$ more than age. However, age and density are correlated as $<200$\,Myr planets planets are known to be puffier than their older counterparts \citep{Vach2024}.
} \label{fig:overview}
\end{figure}

Our findings are consistent with the findings of \citep{orell-miquel_mopys_2024}, who suggest that 0.1-1\,Gyr systems are no more likely to show H$\alpha$ than older planets. A sample size of three is too small to add much, but this does suggest focusing on the youngest systems ($<$50\,Myr) might yield more detections. Although V1298 Tau\,c \citep[$28\pm$4\,Myr;][]{johnson_aligned_2022} showed no evidence of excess H$\alpha$ absorption in transit \citep{feinstein_h-alpha_2021}. Instead, the deciding factor might simply be low density and higher equilibrium temperature, as suggested by Figure~\ref{fig:overview}. If so, the youngest planets may still be favorable as they are known to be larger (lower density) than their older counterparts \citep{mann_zodiacal_2016,Vach2024}.

Fortunately, the number of $<$50\,Myr transiting planetary systems has been rapidly expanding over the last few years. Aside from the two studied here, there are 11 such systems; Kepler-1643, KOI-7913, \& KOI-7368 \citep[40\,Myr;][]{Bouma2022}, TOI-837 \citep[35\,Myr;][]{Barragan2024}, NGTS-33 \citep[30\,Myr;][]{Alves2025}, V1298 Tau \citep[23\,Myr;][]{david_four_2019}, AU Mic \citep[22\,Myr;][]{plavchan_planet_2020}, TOI-1227 \citep[11\,Myr;][]{mann_tess_2022}, K2-33 \citep[10\,Myr;][]{Thao2023}, and IRAS 04125+2902 \citep[3\,Myr;][]{Barber2024b}. \\

Another thing that sets HIP\,67522\,b apart is its low density \citep{Thao2024b}, which makes it ideal for active mass-loss. This would suggest focusing on planets with radii $>6R_\oplus$; TOI-837\,b, TOI-1227\,b, K2-33\,b, NGTS-33\,b, HIP\,67522\,c, IRAS 04125+2902, and V1298\,Tau\,de. These systems are also advantageous because their (likely) low densities make them favorable to detect features of the planet's atmosphere in the same spectra used to gather H$\alpha$ or He~I observations.

\section{Acknowledgments}.

The authors wish to thank Pibby and Bandit for their thoughtful contributions. A.W.M. was supported by a grant from NASA's exoplanet research program (80NSSC25K7148) and from NSF's CAREER program (AST-2143763). R.P.M. was supported by NC Space Grant, a grant from NASA's TESS Guest Investigator program (80NSSC24K0880). 

This work makes use of observations from the LCOGT network. Based on observations made with the Italian Telescopio Nazionale Galileo (TNG) operated by the Fundación Galileo Galilei (FGG) of the Istituto Nazionale di Astrofisica (INAF) at the Observatorio del Roque de los Muchachos (La Palma, Canary Islands, Spain). This research has used data from the CTIO/SMARTS 1.5 m telescope, which is operated as part of the SMARTS Consortium by RECONS members Todd Henry, Hodari James, Wei-Chun Jao, and Leonardo Paredes. 

\vspace{5mm}
\facilities{SMARTS 1.5m (CHIRON)\citep{tokovinin_chironfiber_2013}, TNG (HARPS-N)\citep{cosentino_harps-n_2012}, Magellan (PFS).}

\software{Astropy \citep{robitaille_astropy_2013},
Matplotlib \citep{hunter_matplotlib_2007},
TelFit \citep{gullikson_telfit_2014},
NumPy \citep{harris_array_2020}, 
Specutils \citep{earl_astropyspecutils_2024}, 
\texttt{LBLRTM} \citep{clough_lblrtm_2014}. }

\bibliography{mann_ref.bib,exoplanets_refs.bib}{}
\bibliographystyle{aasjournal}

\end{document}